\documentclass[onecolumn, ceqn, showpacs, showkeys, prd, onecolumn, byrevtex, nofootinbib, amsmath, amsfonts, amssymb]{revtex4-1}

\usepackage{amssymb,amsmath,mathtools,xcolor,graphicx,xspace,colortbl,ragged2e,rotating} %
\usepackage{textcomp}
\usepackage{amsfonts}  
\usepackage{amsmath}  
\usepackage{amssymb}  
\usepackage{boxedminipage}  
\usepackage{graphics}  
\usepackage{graphicx}  
\usepackage{ragged2e}  
\usepackage{tabulary}  
\usepackage{varioref}  
\usepackage{wrapfig}  
\usepackage{xcolor}  
\graphicspath{{FRACTIONAL5_2_graphics/}{FRACTIONAL5_2_tcache/}{FRACTIONAL5_2_gcache/}}
\DeclareGraphicsExtensions{.pdf,.eps,.ps,.png,.jpg,.jpeg}

\begin{document}\thinspace

\title{RELATIVISTIC FRACTIONAL-DIMENSION GRAVITY}
\author{Gabriele U. Varieschi
}
\email[E-mail me at: ]{gvarieschi@lmu.edu
}
\homepage[Visit: ]{http://gvarieschi.lmu.build
}
\affiliation{Department of Physics, Loyola Marymount University, Los Angeles, CA 90045, USA
}

\date{
\today
}
\begin{abstract}
This paper presents a relativistic version of Newtonian Fractional-Dimension Gravity (NFDG), an alternative gravitational model recently introduced and based on the theory of fractional-dimension spaces. This extended version - Relativistic Fractional-Dimension Gravity (RFDG) - is based on other existing theories in the literature and might be useful for astrophysical and cosmological applications.

In particular, in this work we review the mathematical theory for spaces with non-integer dimensions and its connections with the non-relativistic NFDG. The Euler-Lagrange equations for scalar fields can also be extended to spaces with fractional dimensions, by adding an appropriate weight factor, and then can be used to generalize the Laplacian operator for rectangular, spherical, and cylindrical coordinates. In addition, the same weight factor can be added to the standard Hilbert action in order to obtain the field equations, following methods used for scalar-tensor models of gravity, multi-scale spacetimes, and fractional gravity theories.

We then apply the field equations to standard cosmology and to the Friedmann-Lema{\^\i}tre-Robertson-Walker metric. Using a suitable weight $v_{t}\left (t\right )$, depending on the synchronous time $t$ and on a single time-dimension parameter $\alpha _{t}$, we extend the Friedmann equations to the RFDG case. This allows for the computation of the scale factor $a\left (t\right )$ for different values of the fractional time-dimension $\alpha _{t}$  and the comparison with standard cosmology results. Future additional work on the subject, including studies of the cosmological late-time acceleration, type Ia supernovae data, and related dark energy theory will be needed to establish this model as a relativistic alternative theory of gravity.
\end{abstract}
\keywords{fractional-dimension gravity; modified gravity; dark matter; dark energy; cosmology
}
\maketitle

\section{\label{sect:intro}
Introduction
}
 This paper considers a possible relativistic generalization of Newtonian Fractional-Dimension Gravity (NFDG), which was previously introduced as a non-relativistic alternative gravity model (\cite{Varieschi:2020ioh,Varieschi:2020dnd,2021MNRAS.503.1915V}, papers I, II, and III in the following). The main goal of NFDG was to model galactic rotation curves without using the controversial Dark Matter (DM) component (see also \cite{Varieschi:webpage} for a general overview of NFDG). This was done by assuming that galactic structures might behave like fractal media, with an effective spatial dimension which could be lower than the standard value $D =3$, including possible fractional, i.e., non-integer values. Each galaxy was then characterized by a particular form of this varying dimension $D =D\left (r\right )$, which can be a function of the radial distance from the galactic center.

In paper I \cite{Varieschi:2020ioh}, it was shown how NFDG is a natural extension of standard Newtonian gravity to non-integer dimension spaces. Starting from a heuristic extension of Gauss's law for gravity to fractional-dimension spaces, we were able to generalize the gravitational field and potential for extended mass sources, the Laplace and Poisson equations, the multipole expansion, etc. Additionally, we modeled several types of spherically-symmetric galactic structures, such as Plummer models and others, showing that flat rotation curves can be obtained in NFDG without resorting to DM. 

In paper II \cite{Varieschi:2020dnd}, this analysis was extended to axially-symmetric structures in order to model real galactic data from the Spitzer Photometry and Accurate Rotation Curves (SPARC) database \cite{Lelli:2016zqa}. In addition to exponential, Kuzmin, and other similar thin/thick disk mass distributions, the case of the disk-dominated dwarf spiral galaxy NGC 6503 was considered and it was shown that the rotation curve of this galaxy could be obtained by simply assuming $D\left (r\right ) \approx 2$ over most of the radial range. In other words, if this galaxy were actually to behave like a fractal medium with (Hausdorff) dimension $D\left (r\right ) \approx 2$, its dynamics would be fully explained by NFDG without any DM\ contribution.

In addition to NGC 6503, in paper III \cite{2021MNRAS.503.1915V} we studied two additional galaxies with our methods: NGC 7814 (bulge-dominated spiral) and NGC 3741 (gas-dominated dwarf). Although these two galaxies seem to be characterized by different functions for the varying dimension $D =D\left (r\right )$, their rotation curves were also fully fitted with NFDG methods, again without any DM. In paper III, the use of a variable dimension $D\left (r\right )$ as a function of the field point was also discussed and justified in terms of other similar existing studies. In all these three papers, we pointed out that NFDG is only loosely based on the methods of fractional calculus and fractional mechanics (see \cite{Varieschi:2018} and references therein), but is not a fractional theory in the sense used by other gravitational models \cite{Calcagni:2009kc,Calcagni:2010bj,Calcagni:2011kn,Calcagni:2011sz,Calcagni:2013yqa,Calcagni:2016azd,Calcagni:2016xtk,Calcagni:2018dhp,Calcagni:2020ads,Calcagni:2021ljs,Calcagni:2021ipd,Calcagni:2021aap,Giusti:2020rul,Giusti:2020kcv}. NFDG field equations are of integer order, therefore local, as opposed to non-local field equations based on fractional differential operators.

NFDG also shows possible connections with Modified Newtonian Dynamics
(MOND) \cite{Milgrom:1983ca,Milgrom:1983pn,Milgrom:1983zz}, as discussed in detail in our previous papers \cite{Varieschi:2020ioh,Varieschi:2020dnd,2021MNRAS.503.1915V}. In particular, MOND phenomenology, including the recently reported Radial Acceleration Relation (RAR) \cite{McGaugh:2016leg,Lelli:2017vgz,Chae:2020omu}, might be explained by our varying dimension $D =D\left (r\right )$, which provides the link between the inherently non-linear MOND theory and the linear NFDG.

In this paper, we focus our efforts instead on a relativistic version of our model, which will be called Relativistic Fractional-Dimension Gravity (RFDG). This extended version of NFDG is very similar to Calcagni's theory with ordinary derivatives \cite{Calcagni:2010bj,Calcagni:2013yqa}, but uses the weight factors introduced in our previous papers I-III. Other more limited analyses of relativistic equations for non-integer dimension spaces are found in the literature \cite{2006CzJPh..56..323S,Sadallah:2009zz}, but they don't fully explore the subject. These relativistic approaches to non-integer, lower-dimension spaces should not be confused with past attempts to study General Relativity (GR) in two or three-dimensional spacetimes \cite{1977AmJPh..45..833C,Romero:1994va,1984AnPhy.152..220D}: in NFDG (or RFDG) the spacetime is the usual $3 +1$, while we consider possible subsets $X \subset \mathbb{R}^{3}$ of the standard tri-dimensional space, whose Hausdorff dimensions can be $D \neq 3$, and possibly also a fractional time dimension in RFDG.

Our RFDG model follows the lines of the many existing modified gravity theories in the literature (see \cite{CLIFTON20121} for a general review, or the more recent Ref. \cite{CANTATA:2021ktz}) and their possible cosmological consequences. As in standard GR \cite{Will:2014kxa}, an alternative model of gravity should be tested against experimental results of gravitational physics, or at least be consistent with General Relativity at scales where Einstein's theory is undisputed. For example, the MOND model is well established as an alternative gravitational theory (for general reviews see Refs. \cite{Sanders:2002pf,Famaey:2011kh}) and its many implications for gravitation and cosmology have been studied for decades, determining the strong and weak points of the model. On the contrary, our NFDG and RFDG are very recent models with limited results and need to be analyzed in more detail through future work, in order to become viable alternatives to standard GR.

In Sect. \ref{sect:Mathematical}, we will describe the mathematical theory for spaces with fractional dimension and review the fundamental NFDG results from our previous papers. In Sect. \ref{sect::Euler}, we
will review and expand the Euler-Lagrange equations for non-integer dimension spaces, while in Sect. \ref{sect::Relativistic} we will detail the relativistic equations and apply them to standard cosmology. Finally, in
Sect. \ref{sect::conclusion} conclusions are drawn
and possible future work on the subject is outlined.

\section{\label{sect:Mathematical} Mathematical theory for spaces with non-integer dimension and NFDG
}
The dimensions of space and spacetime play an important role in determining the form of the physical laws and of the constants of nature. While we perceive space as three-dimensional (and time as one-dimensional), discussions on a possible explanation of the tri-dimensionality of space date back to Ptolemy and the early Greeks \cite{10.2307/37418}. In modern times, Ehrenfest's famous article on the subject \cite{doi:10.1002/andp.19203660503} explained how the tri-dimensionality of space is inherently connected with fundamental physical laws, such as those of stable planetary orbits, atoms and molecules stability, and several others. More recent discussions about the dimensionality of space can be found in the works by Barrow  \cite{10.2307/37418}, Callender \cite{2005SHPMP..36..113C}, and references therein.

With more recent advances in mathematical theories and fractal geometries, it also became possible to consider a continuous variation in the number of dimensions $D$ for space, i.e., not just positive integer dimensions, but any real (or even complex) spatial dimension $D$. Although this possibility emerged in several areas of physics, it became popular in dimensional regularization techniques commonly used in
quantum field theory \cite{Bollini:1972ui,tHooft:1972tcz,Wilson:1972cf}. As part of these techniques (see also \cite{1995iqft.book.....P}, page
249), the area of a unit hypersphere
$S$ in $D$ dimensions was evaluated as
$\int _{S}d\Omega _{D} =\frac{2\pi ^{D/2}}{\Gamma (D/2)}$, which yields familiar results for integer values of $D$, such as $2$ for $D =1$, $2\pi $ for $D =2$, $4\pi $ for $D =3$, etc.

A more comprehensive theory for spaces with non-integer dimension was first introduced by Stillinger in 1977 \cite{doi:10.1063/1.523395}. Starting from quantities depending explicitly on a variable dimension $D$, such as the Gaussian integral $\int d\mathbf{r}\exp \left ( -\alpha r^{2}\right ) =\left (\pi /\alpha \right )^{D/2}$, or the radial Laplace operator $\frac{1}{r^{D -1}}\frac{d}{dr}\left (r^{D -1}\frac{d}{dr}\right ) =\frac{d^{2}}{dr^{2}} +\frac{\left (D -1\right )}{r}\frac{d}{dr}$, an axiomatic theory for metric spaces of non-integer dimension was then introduced, based on weights, $W_{n}\left (\mathbf{x}_{1} , . . . ,\mathbf{x}_{n}\vert r_{1} , . . . ,r_{n}\right )$, for a fixed set of points $\mathbf{x}_{1} , . . .\mathbf{x}_{n}$ and distances $r_{1} , . . . ,r_{n}$ measured from them. The simplest of these weights, $W_{1}$, was computed as $W_{1}\left (r\right ) =\sigma \left (D\right )r^{D -1} =\frac{2\pi ^{D/2}}{\Gamma \left (D/2\right )}r^{D -1}$, with the radial distance $r$ measured from the origin.

This weight allows for the generalization of the integral of a
spherically-symmetric function
$f =f(r)$
over a
$D$-dimensional metric space as $\int _{0}^{\infty }f(r)W_{1}\left (r\right )dr =\frac{2\pi ^{D/2}}{\Gamma (D/2)}\int _{0}^{\infty }f\left (r\right )r^{D -1}dr$, and of the volume of the radius-R sphere as $V\left (R ,D\right ) =\int _{0}^{R}W_{1}\left (r\right )dr =\frac{\pi ^{D/2}R^{D}}{\Gamma \left (1 +D/2\right )}$.

In the same paper  \cite{doi:10.1063/1.523395}, Stillinger introduced a generalized Laplacian in polar coordinates:\newline  $ \nabla ^{2}g =\left [\frac{1}{r^{D -1}}\frac{ \partial }{ \partial r}\left (r^{D -1}\frac{ \partial }{ \partial r}\right ) +\frac{1}{r^{2}\sin ^{D -2}\theta }\frac{ \partial }{ \partial \theta }\left (\sin ^{D -2}\theta \frac{ \partial }{ \partial \theta }\right )\right ]g =\left [\frac{ \partial ^{2}}{ \partial r^{2}} +\frac{\left (D -1\right )}{r}\frac{ \partial }{ \partial r} +\frac{1}{r^{2}}\left (\frac{ \partial ^{2}}{ \partial \theta ^{2}} +\frac{\left (D -2\right )}{\tan \theta }\frac{ \partial }{ \partial \theta }\right )\right ]g$, and applied it to the solution of the generalized two-dimensional Schr{\"o}dinger's equation, with the angular solution expressed in terms of Gegenbauer polynomials.

As for the physical meaning of a non-integer dimension $D$, Stillinger roughly estimated the possible uncertainty of the spatial dimension as $D \simeq 3 \pm 10^{ -6}$ in our terrestrial locale and also explored the possibility of the role of $D$ as a field variable in geometric theories of gravity. In particular, he stated \cite{doi:10.1063/1.523395}: ``However a more general class of spaces can also be generated within which $D$ varies continuously from point to point (integration weights $W_{n}$ would exhibit the change explicitly)''. This seems to imply that the axiomatic bases for non-integer dimension spaces would still be valid for the weight $W_{1}$ generalized as $W_{1}\left (r\right ) =\sigma \left [\left .D\left (r\right )\right .\right ]r^{D\left (r\right ) -1} =\frac{2\pi ^{D\left (r\right )/2}}{\Gamma \left [D\left (r\right )/2\right ]}r^{D\left (r\right ) -1}$, with $D =D\left (r\right )$ an explicit function of the field point. This assumption was used as the rationale for a varying fractional dimension $D$ in all our three NFDG papers.

A similar but different approach was later introduced by Svozil \cite{1987JPhA...20.3861S}, within the framework of the Hausdorff measure theory. This lead directly to the integral of a
spherically-symmetric function
$f =f(r)$
over a
$D$-dimensional metric space $\chi $ as follows:
\begin{equation}\int _{\chi }fd\mu _{H} =\frac{2\pi ^{D/2}}{\Gamma (D/2)}\int _{0}^{\infty }f(r)r^{D -1}dr , \label{eq2.1}
\end{equation}
where
$\mu _{H}$
denotes an appropriate Hausdorff measure over the space. This result is the same obtained previously by Stillinger and was also connected by Svozil to the Weyl's fractional integral defined as
$W^{ -D}f(x) =\frac{1}{\Gamma (D)}\int _{x}^{\infty }(t -x)^{D -1}f(t)dt$, so that Eq. (\ref{eq2.1}) can also be written as $\int _{\chi }fd\mu _{H} =\frac{2\pi ^{D/2}\Gamma (D)}{\Gamma (D/2)}W^{ -D}f(0)$, thus connecting the theory of non-integer dimension spaces with fractional calculus.

In 2004, Palmer and Stavrinou \cite{Palmer_2004} expanded the previous concepts into the theory of the equations of motion in a non-integer-dimensional space by using Svozil's measure theory approach and multi-variable integration techniques. In particular, to integrate over a subset $X \subset \mathbb{R}^{3}$ they assumed that $X =X_{1} \times X_{2} \times X_{3}$, where each metric space $X_{i}$
($i =1 ,2 ,3$) is equipped with a Hausdorff measure
$\mu _{i}(X_{i})$
and a dimension
$\alpha _{i}$. When $\alpha _{i} =1$, the Hausdorff measure simply becomes a Lebesgue measure. The Hausdorff measure for the product set $X$
can be defined as
$\mu _{H}(X) =(\mu _{1} \times \mu _{2} \times \mu _{3})(X_{1} \times X_{2} \times X_{3}) =\mu _{1}(X_{1})\mu _{2}(X_{2})\mu _{3}(X_{3})$
and the overall Hausdorff spatial dimension is then
$D =\alpha _{1} +\alpha _{2} +\alpha _{3}$.

 Applying Fubini's theorem we have \cite{Palmer_2004,bookTarasov,bookZubair,Tarasov:2014fda,TARASOV2015360}:

\begin{gather}\int _{X}f(x_{1} ,x_{2} ,x_{3})d\mu _{H} =\int _{X_{1}}\int _{X_{2}}\int _{X_{3}}f(x_{1} ,x_{2} ,x_{3})d\mu _{1}(x_{1})d\mu _{2}(x_{2})d\mu _{3}(x_{3}) , \label{eq2.2} \\
d\mu _{i}(x_{i}) =\frac{\pi ^{\alpha _{i}/2}}{\Gamma (\alpha _{i}/2)}\left \vert x_{i}\right \vert ^{\alpha _{i} -1}dx_{i} ,\ i =1 ,2 ,3 \nonumber \end{gather}where the infinitesimal measures $d\mu _{i}$ in the second line of the previous equations follow from the original Stillinger's weight $W_{1}$ described above, and used in the integral in Eq. (\ref{eq2.1}). The factor of two in the weight $W_{1}$ is now omitted, assuming integration between $ -\infty $ and $ +\infty $ in each sub-space $X_{i}$. 

As was noted in paper I, it is easy to check that the integral in Eq. (\ref{eq2.2}) when applied to a function
$f(x_{1} ,x_{2} ,x_{3}) =f(r)$ in spherical coordinates
$(r ,\theta  ,\varphi )$, yields the expression in Eq. (\ref{eq2.1}). This follows from the
standard relations between rectangular and spherical coordinates and from the definitions
for the differential measures in the second line of Eq. (\ref{eq2.2}):
$d\mu _{1}d\mu _{2}d\mu _{3} =\frac{\pi ^{\alpha _{1}/2}}{\Gamma \left (\alpha _{1}/2\right )}\frac{\pi ^{\alpha _{2}/2}}{\Gamma \left (\alpha _{2}/2\right )}\frac{\pi ^{\alpha _{3}/2}}{\Gamma \left (\alpha _{3}/2\right )}r^{\alpha _{1} +\alpha _{2} +\alpha _{3} -1}dr\vert \sin \theta \vert ^{\alpha _{1} +\alpha _{2} -1}\left \vert \cos \theta \right \vert ^{\alpha _{3} -1}d\theta \vert \sin \varphi \vert ^{\alpha _{2} -1}\left \vert \cos \varphi \right \vert ^{\alpha _{1} -1}d\varphi $. Performing the angular integrations, simplifying the results, and using
$D =\alpha _{1} +\alpha _{2} +\alpha _{3}$, the result in Eq. (\ref{eq2.1}) is readily obtained.

While this result is independent of how the dimensions $\alpha _{i}$ arrange themselves to act on the orthogonal coordinates and depends only on the overall dimension $D$,  Palmer and Stavrinou \cite{Palmer_2004} also noted that in more general cases it is not clear if the non-integer dimension $D$ distributes itself over the $n$ space coordinates (example: $\alpha _{1} =\alpha _{2} = . . . =\alpha _{n} =D/n$) or on only one coordinate (example: $\alpha _{1} =\alpha _{2} = . . . =\alpha _{n -1} =1$ and $\alpha _{n} =D -\left (n -1\right )$), eventually favoring the latter case in Ref. \cite{Palmer_2004}. In these more general cases, the results of the integrations in Eq. (\ref{eq2.2}) will depend on how this choice for the $\alpha _{i}$ dimensions is made.

With all these assumptions, NFDG was developed in papers I-III \cite{Varieschi:2020ioh,Varieschi:2020dnd,2021MNRAS.503.1915V} by extending Gauss's law for gravitation to lower-dimensional spacetime $D +1$, with non-integer space dimension $0 <D \leq 3$. A scale length $l_{0}$ was needed for dimensional correctness of all expressions when $D \neq 3$, so that dimensionless coordinates were adopted in all formulas, such as the rescaled radial distance $w_{r} \equiv r/l_{0}$ or, in general, the dimensionless coordinates
$\mathbf{w} \equiv \mathbf{x}/l_{0}$
for the field point and
$\mathbf{w}^{ \prime } \equiv \mathbf{x}^{ \prime }/l_{0}$
for the source point. A rescaled mass ``density'' was also introduced: $\widetilde{\rho }\left (\mathbf{w}^{ \prime }\right ) =\rho \left (\mathbf{w}\mathbf{}^{ \prime }l_{0}\right )l_{0}^{3} =\rho \left (\mathbf{x}^{ \prime }\right )l_{0}^{3}$, where
$\rho (\mathbf{x}^{ \prime })$
is the standard mass density in
$\mbox{kg}\thinspace \mbox{m}^{ -3}$, and with $d\widetilde{m}_{\left (D\right )} =\widetilde{\rho }\left (\mathbf{w}^{ \prime }\right )d^{D}\mathbf{w}^{ \prime }$ representing the infinitesimal source mass in a D-dimensional space.\protect\footnote{
SI units will be used throughout this paper, unless otherwise noted.
}

The NFDG gravitational potential
$\widetilde{\phi }\left (\mathbf{w}\right )$
was then obtained as:
\begin{gather}\widetilde{\phi }(\mathbf{w}) = -\frac{2\pi ^{1 -D/2}\Gamma (D/2)G}{\left (D -2\right )l_{0}}{\displaystyle\int _{V_{D}}}\frac{\widetilde{\rho }(\mathbf{w}^{ \prime })}{\left \vert \mathbf{w} -\mathbf{w}^{ \prime }\right .\vert ^{D -2}}d^{D}\mathbf{w}^{ \prime };\ D \neq 2 \label{eq2.3} \\
\widetilde{\phi }\left (\mathbf{w}\right ) =\frac{2G}{l_{0}}{\displaystyle\int _{V_{2}}}\widetilde{\rho }\left (\mathbf{w}^{ \prime }\right )\ln \left \vert \mathbf{w} -\mathbf{w}^{ \prime }\right .\vert d^{2}\mathbf{w}^{ \prime };\ D =2 \nonumber \end{gather}
where  the physical dimensions for the NFDG\ gravitational potential
$\widetilde{\phi }$ are the same as those of the standard Newtonian potential (i.e., measured in $\mbox{m}^{2}\thinspace \mbox{s}^{ -2}$).

Assuming that
$\widetilde{\phi }(\mathbf{w})$
and the NFDG gravitational field
$\mathbf{g}(\mathbf{w})$
are connected by
$\mathbf{g}(\mathbf{w}) = - \nabla _{D}\widetilde{\phi }(\mathbf{w})/l_{0}$, where the D-dimensional gradient
$ \nabla _{D}$
is equivalent to the standard one, but derivatives are taken with respect to the rescaled coordinates $\mathbf{w}$, we also obtained:
\begin{equation}\mathbf{g}(\mathbf{w}) = -\frac{2\pi ^{1 -D/2}\Gamma (D/2)G}{l_{0}^{2}}{\displaystyle\int _{V_{D}}}\widetilde{\rho }(\mathbf{w}^{ \prime })\frac{\mathbf{w} -\mathbf{w}^{ \prime }}{\left \vert \mathbf{w} -\mathbf{w}^{ \prime }\right .\vert ^{D}}d^{D}\mathbf{w}^{ \prime } . \label{eq2.4}
\end{equation}
It is easy to check that the expressions in Eqs. (\ref{eq2.3})-(\ref{eq2.4}) above correctly reduce to the standard Newtonian ones for $D =3$. The gravitational potential and field in the last two equations were derived for a fixed value of the fractional dimension $D$, but it was argued that they could also be applicable to the case of a variable dimension $D\left (\mathbf{w}\right )$, assuming a slow change of this dimension with the field point coordinates.

The scale length $l_{0}$ was related to the MOND\  acceleration constant
$a_{0}$ (sometimes also denoted by $g_{\dag }$ \cite{McGaugh:2016leg,Lelli:2017vgz}):

\begin{equation}a_{0} \equiv g_{\dag } =1.20 \pm 0.02\ \text{(random)} \pm 0.24\ \text{(syst)} \times 10^{ -10}\ \mbox{}\ \mbox{m}\thinspace \mbox{s}^{ -2} , \label{eq2.5}
\end{equation}
which represents the acceleration scale below which MOND corrections are needed. In papers I-III, a possible connection between the scale length $l_{0}$ and the MOND\ acceleration $a_{0}$ was proposed as:

\begin{equation}a_{0} \approx \frac{GM}{l_{0}^{2}} , \label{eq2.6}
\end{equation}where $M$ is the mass of the system being studied (or a suitable reference mass). The main consequences of the MOND\ theory (the flat rotation velocity
$V_{f} \approx \sqrt[{4}]{GMa_{0}}$, the ``baryonic'' Tully-Fisher relation-BTFR
$M_{bar} \sim V_{f}^{4}$, etc.) were recovered in NFDG by considering the MOND limit to be equivalent to a space dimension $D \approx 2$  \cite{Varieschi:2020ioh}.

The main NFDG equations  (\ref{eq2.3})-(\ref{eq2.4}) were then adapted to spherically-symmetric and axially-symmetric cases of interest, then leading to detailed fits of galactic rotation curves for three notable cases (NGC 6503, NGC 7814, NGC 3741) as outlined in Sect. \ref{sect:intro} above. It should be noted that the integrations over $D$-dimensional spaces were performed following the techniques based on Eqs. (\ref{eq2.1})-(\ref{eq2.2}) and for different choices of how the individual dimensions $\alpha _{1}$, $\alpha _{2}$, $\alpha _{3}$ arrange themselves on the three spatial orthogonal coordinates (see papers I-III for full details).

\section{\label{sect::Euler} Euler-Lagrange equations for spaces with non-integer dimension
}
In this section, we will expand the treatment of the Euler-Lagrange equations for fields in non-integer-dimension spaces introduced by Palmer and Stavrinou \cite{Palmer_2004}, and use it as a starting point for the relativistic equations of motion. This approach has the obvious advantage of yielding the dynamics of the field for any number of degrees of freedom and in any coordinate basis.

We assume a Lagrangian density in four spacetime coordinates, $\mathcal{L} =\mathcal{L}\left (\phi  , \partial _{\mu }\phi \right )$, where the field $\phi $ and $ \partial _{\mu }\phi $ are functions of $\left (t ,x^{1} ,x^{2} ,x^{3}\right )$ and with $ \partial _{\mu } =\left ( \partial _{t} , \partial _{x^{1}} , \partial _{x^{2}} , \partial _{x^{3}}\right )$. The generalized action $S$ in a $D +1$ spacetime is \cite{Palmer_2004}:\protect\footnote{
In Sect. \ref{sect::Relativistic}, we will also include a possible time weight $v_{t}\left (t\right ) =\frac{\pi ^{\alpha _{t}/2}}{\Gamma \left (\alpha _{t}/2\right )}\left \vert t\right \vert ^{\alpha _{t} -1}$ into the action. Following the original analysis in Ref. \cite{Palmer_2004}, we will not use this weight in this section.
}

\begin{gather}S =\int dtd^{D}x\mathcal{L}\left (\phi  , \partial _{\mu }\phi \right ) =\int dt\int d\mu _{1}\left (x^{1}\right )d\mu _{2}\left (x^{2}\right )d\mu _{3}\left (x^{3}\right )\mathcal{L}\left (\phi  , \partial _{\mu }\phi \right ) \label{eq3.1} \\
d\mu _{i}\left (x^{i}\right ) =W_{1}\left (x^{i} ,\alpha _{i}\right )dx^{i} =\frac{\pi ^{\alpha _{i}/2}}{\Gamma (\alpha _{i}/2)}\left \vert x^{i}\right \vert ^{\alpha _{i} -1}dx^{i} ,\ i =1 ,2 ,3 \nonumber \end{gather}where the measures $d\mu _{i}$ are those from Eq. (\ref{eq2.2}) and all the integrations now extend from $ -\infty $ to $ +\infty $, so that the measure weights are $W_{1}\left (x^{i} ,\alpha _{i}\right ) =\frac{\pi ^{\alpha _{i}/2}}{\Gamma \left (\alpha _{i}/2\right )}\left \vert x^{i}\right \vert ^{\alpha _{i} -1}$ (the factor of two in the original Stillinger's weight is now omitted).\protect\footnote{
Dimensionless coordinates, such as $w^{i} =x^{i}/l_{0}$, $w_{r} =r/l_{0}$, etc., should be used in most equations in this section and in the following ones. For simplicity's sake, in this paper we have left standard coordinates ($x^{i}$, $r$, $R$, etc.) in most equations, without transforming them into dimensionless, rescaled ones.
}

By taking variations and minimizing the action \cite{Palmer_2004}, it is straightforward to obtain the following Euler-Lagrange equations:

\begin{gather}\prod \limits _{i =1}^{3}W_{1}\left (x^{i} ,\alpha _{i}\right )\frac{ \partial \mathcal{L}\left (\phi  , \partial _{\mu }\phi \right )}{ \partial \phi } -\prod \limits _{i =1}^{3}W_{1}\left (x^{i} ,\alpha _{i}\right ) \partial _{\mu }\frac{ \partial \mathcal{L}\left (\phi  , \partial _{\mu }\phi \right )}{ \partial \left ( \partial _{\mu }\phi \right )} -\frac{ \partial \mathcal{L}\left (\phi  , \partial _{\mu }\phi \right )}{ \partial \left ( \partial _{\mu }\phi \right )} \partial _{\mu }\prod \limits _{i =1}^{3}W_{1}\left (x^{i} ,\alpha _{i}\right ) \label{eq3.2} \\
 =\prod \limits _{i =1}^{3}W_{1}\left (x^{i} ,\alpha _{i}\right )\frac{ \partial \mathcal{L}\left (\phi  , \partial _{\mu }\phi \right )}{ \partial \phi } - \partial _{\mu }\left [\prod \limits _{i =1}^{3}W_{1}\left (x^{i} ,\alpha _{i}\right )\frac{ \partial \mathcal{L}\left (\phi  , \partial _{\mu }\phi \right )}{ \partial \left ( \partial _{\mu }\phi \right )}\right ] =0 , \nonumber \end{gather}with the measure weights $W_{1}\left (x^{i} ,\alpha _{i}\right )$ described above, or even for more general types of measures. Since for $D =3$, and $\alpha _{1} =\alpha _{2} =\alpha _{3} =1$, we have $\prod \limits _{i =1}^{3}W_{1}\left (x^{i} ,\alpha _{i}\right ) =1$ and $ \partial _{\mu }\prod \limits _{i =1}^{3}W_{1}\left (x^{i} ,\alpha _{i}\right ) =0$, Eq. (\ref{eq3.2}) reduces to standard Euler-Lagrange equations in $3 +1$ spacetimes.

As noted in Ref. \cite{Palmer_2004}, the ``flow'' or ``current'' of the measure $ \partial _{\mu }\prod \limits _{i =1}^{3}W_{1}\left (x^{i} ,\alpha _{i}\right )$, multiplied by the momentum density of the field $\frac{ \partial \mathcal{L}\left (\phi  , \partial _{\mu }\phi \right )}{ \partial \left ( \partial _{\mu }\phi \right )}$ in the third term of the first line in Eq. (\ref{eq3.2}), will alter the dynamics of the field $\phi $ in a non-integer-dimensional space, compared to the standard case. As a consequence, if the system is invariant under a symmetry transformation $\phi \left (x\right ) \rightarrow \phi \left (x\right ) +\delta \phi \left (x\right )$, the related conserved current density and conservation law in non-integer dimensions are \cite{Palmer_2004}:

\begin{gather}J^{\mu } =\prod \limits _{i =1}^{3}W_{1}\left (x^{i} ,\alpha _{i}\right )\frac{ \partial \mathcal{L}\left (\phi  , \partial _{\mu }\phi \right )}{ \partial \left ( \partial _{\mu }\phi \right )}\delta \phi  \label{eq3.3} \\
 \partial _{\mu }J^{\mu } =0. \nonumber \end{gather}

This last equation, and the previous Eq. (\ref{eq3.2}), could have been also introduced from the standard equations by substituting $\mathcal{L} \rightarrow \prod \limits _{i =1}^{3}W_{1}\left (x^{i} ,\alpha _{i}\right )\mathcal{L}$ and $J^{\mu } \rightarrow \prod \limits _{i =1}^{3}W_{1}\left (x^{i} ,\alpha _{i}\right )J^{\mu }$, respectively. To conclude this general overview, we will outline in the following sub-sections the specific cases of rectangular, spherical, and cylindrical coordinates and the related $D$-dimensional Laplace operators.

\subsection{\label{subsect:rectangular}Rectangular coordinates}
In rectangular coordinates, the generalized Euler-Lagrange equations can be obtained directly from Eq. (\ref{eq3.2}) with the weights in Eq. (\ref{eq3.1}) \cite{Palmer_2004}:

\begin{equation}\frac{ \partial \mathcal{L}\left (\phi  , \partial _{\mu }\phi \right )}{ \partial \phi } - \partial _{\mu }\frac{ \partial \mathcal{L}\left (\phi  , \partial _{\mu }\phi \right )}{ \partial \left ( \partial _{\mu }\phi \right )} -\left (\alpha _{\mu \nu } -\delta _{\mu \nu }\right )\left (x^{\left ( -1\right )}\right )^{\nu }\frac{ \partial \mathcal{L}\left (\phi  , \partial _{\mu }\phi \right )}{ \partial \left ( \partial _{\mu }\phi \right )} =0 , \label{eq3.4}
\end{equation}where $\alpha _{\mu \nu } =diag\left (1 ,\alpha _{1} ,\alpha _{2} ,\alpha _{3}\right )$, $\delta _{\mu \upsilon }$ is the diagonal unit matrix, $x^{\left ( -1\right )} =column\left (t^{ -1} ,\left (x^{1}\right )^{ -1} ,\left (x^{2}\right )^{ -1} ,\left (x^{3}\right )^{ -1}\right )$, with $\mu  ,\nu  =0 ,1 ,2 ,3$. The total spacetime dimension is $D_{t} =1 +D =1 +\alpha _{1} +\alpha _{2} +\alpha _{3} =Tr\left (\alpha _{\mu \nu }\right )$, where the time dimension is assumed to be integer.

As in the original treatment for the Schr{\"o}dinger's equation \cite{Palmer_2004,morse1953methods}, we can consider $\phi $ and $\phi ^{ \ast }$ as separate fields which can be varied independently and then use the Lagrangian density $\mathcal{L} = \nabla \phi ^{ \ast } \cdot  \nabla \phi  = \partial _{i}\phi ^{ \ast } \partial _{i}\phi $ to obtain the generalized Laplace equation, using Eq. (\ref{eq3.4}) for the ``mirror'' field $\phi ^{ \ast }$. The Laplace equation becomes $ \bigtriangledown _{\alpha _{1} ,\alpha _{2} ,\alpha _{3}}^{2}\phi \left (x ,y ,z\right ) =0$, where the generalized Laplacian operator written in standard rectangular coordinates $x$, $y$, $z$, is:

\begin{gather} \bigtriangledown _{\alpha _{1} ,\alpha _{2} ,\alpha _{3}}^{2}\phi \left (x ,y ,z\right ) =\left [\frac{1}{x^{\alpha _{1} -1}}\frac{ \partial }{ \partial x}\left (x^{\alpha _{1} -1}\frac{ \partial }{ \partial x}\right ) +\frac{1}{y^{\alpha _{2} -1}}\frac{ \partial }{ \partial y}\left (y^{\alpha _{2} -1}\frac{ \partial }{ \partial y}\right ) +\frac{1}{z^{\alpha _{3} -1}}\frac{ \partial }{ \partial z}\left (z^{\alpha _{3} -1}\frac{ \partial }{ \partial z}\right )\right ]\phi  \label{eq3.5} \\
 =\left [\frac{ \partial ^{2}}{ \partial x^{2}} +\frac{\left (\alpha _{1} -1\right )}{x}\frac{ \partial }{ \partial x} +\frac{ \partial ^{2}}{ \partial y^{2}} +\frac{\left (\alpha _{2} -1\right )}{y}\frac{ \partial }{ \partial y} +\frac{ \partial ^{2}}{ \partial z^{2}} +\frac{\left (\alpha _{3} -1\right )}{z}\frac{ \partial }{ \partial z}\right ]\phi  . \nonumber \end{gather}The non-integer dimension can then be assigned to just one of the three coordinates  (example: $\alpha _{1} =\alpha _{2} =1$ and $\alpha _{3} =D -2$), or distributed over the three coordinates (example: $\alpha _{1} =\alpha _{2} =\alpha _{3} =D/3$).

\subsection{\label{subsect:spherical}Spherical coordinates
}
To obtain similar results in spherical coordinates $r$, $\theta $, $\varphi $, we could transform directly Eqs. (\ref{eq3.4})-(\ref{eq3.5}), or use the orthonormal basis $ \partial _{\mu } =\left (\frac{ \partial }{ \partial t} ,\frac{ \partial }{ \partial r} ,\frac{1}{r}\frac{ \partial }{ \partial \theta } ,\frac{1}{r\sin \theta }\frac{ \partial }{ \partial \varphi }\right )$. Following this latter option and using again a Lagrangian density $\mathcal{L} = \nabla \phi ^{ \ast } \cdot  \nabla \phi  = \partial _{i}\phi ^{ \ast } \partial _{i}\phi $ in Eq. (\ref{eq3.2}), we obtain:

\begin{gather} \nabla _{\alpha _{1} ,\alpha _{2} ,\alpha _{3}}^{2}\phi \left (r ,\theta  ,\varphi \right ) =\left [\frac{ \partial ^{2}\phi }{ \partial r^{2}} +\frac{\left (\alpha _{1} +\alpha _{2} +\alpha _{3} -1\right )}{r}\frac{ \partial \phi }{ \partial r}\right ] \label{eq3.6} \\
 +\frac{1}{r^{2}}\left [\frac{ \partial ^{2}\phi }{ \partial \theta ^{2}} +\frac{\left (\alpha _{1} +\alpha _{2} -1\right )}{\tan \theta }\frac{ \partial \phi }{ \partial \theta } +\frac{\left (1 -\alpha _{3}\right )}{\cot \theta }\frac{ \partial \phi }{ \partial \theta }\right ] +\frac{1}{r^{2}\sin ^{2}\theta }\left [\frac{ \partial ^{2}\phi }{ \partial \varphi ^{2}} +\frac{\left (\alpha _{2} -1\right )}{\tan \varphi }\frac{ \partial \phi }{ \partial \varphi } +\frac{\left (1 -\alpha _{1}\right )}{\cot \varphi }\frac{ \partial \phi }{ \partial \varphi }\right ] . \nonumber \end{gather}The previous equation extends the results in Ref. \cite{Palmer_2004}, by providing the most general spherical Laplacian for $D =\alpha _{1} +\alpha _{2} +\alpha _{3}$ ($0 <\alpha _{1} ,\alpha _{2} ,\alpha _{3} \leq 1$). For $\alpha _{1} =\alpha _{2} =\alpha _{3} =1$ ($D =3$), the standard spherical Laplacian is recovered, while special cases are obtained if the non-integer dimension is assigned to just one of the three parameters.

If the non-integer parameter is the first one, that is $0 <\alpha _{1} <1$, $\alpha _{2} =\alpha _{3} =1$, $D =\alpha _{1} +2$, we have:
\begin{gather} \nabla _{D -2 ,1 ,1}^{2}\phi \left (r ,\theta  ,\varphi \right ) =\left [\frac{ \partial ^{2}\phi }{ \partial r^{2}} +\frac{\left (D -1\right )}{r}\frac{ \partial \phi }{ \partial r}\right ] \label{eq3.7} \\
 +\frac{1}{r^{2}}\left [\frac{ \partial ^{2}\phi }{ \partial \theta ^{2}} +\frac{\left (D -2\right )}{\tan \theta }\frac{ \partial \phi }{ \partial \theta }\right ] +\frac{1}{r^{2}\sin ^{2}\theta }\left [\frac{ \partial ^{2}\phi }{ \partial \varphi ^{2}} +\frac{\left (3 -D\right )}{\cot \varphi }\frac{ \partial \phi }{ \partial \varphi }\right ] . \nonumber \end{gather}If instead, $0 <\alpha _{2} <1$, $\alpha _{1} =\alpha _{3} =1$, $D =\alpha _{2} +2$, we have:
\begin{gather} \nabla _{1 ,D -2 ,1}^{2}\phi \left (r ,\theta  ,\varphi \right ) =\left [\frac{ \partial ^{2}\phi }{ \partial r^{2}} +\frac{\left (D -1\right )}{r}\frac{ \partial \phi }{ \partial r}\right ] \label{eq3.8} \\
 +\frac{1}{r^{2}}\left [\frac{ \partial ^{2}\phi }{ \partial \theta ^{2}} +\frac{\left (D -2\right )}{\tan \theta }\frac{ \partial \phi }{ \partial \theta }\right ] +\frac{1}{r^{2}\sin ^{2}\theta }\left [\frac{ \partial ^{2}\phi }{ \partial \varphi ^{2}} +\frac{\left (D -3\right )}{\tan \varphi }\frac{ \partial \phi }{ \partial \varphi }\right ] . \nonumber \end{gather}Finally, if $0 <\alpha _{3} <1$, $\alpha _{1} =\alpha _{2} =1$, $D =\alpha _{3} +2$,  we obtain:
\begin{gather} \nabla _{1 ,1 ,D -2}^{2}\phi \left (r ,\theta  ,\varphi \right ) =\left [\frac{ \partial ^{2}\phi }{ \partial r^{2}} +\frac{\left (D -1\right )}{r}\frac{ \partial \phi }{ \partial r}\right ] \label{eq3.9} \\
 +\frac{1}{r^{2}}\left [\frac{ \partial ^{2}\phi }{ \partial \theta ^{2}} +\frac{1}{\tan \theta }\frac{ \partial \phi }{ \partial \theta } +\frac{\left (3 -D\right )}{\cot \theta }\frac{ \partial \phi }{ \partial \theta }\right ] +\frac{1}{r^{2}\sin ^{2}\theta }\left [\frac{ \partial ^{2}\phi }{ \partial \varphi ^{2}}\right ] . \nonumber \end{gather}

In Ref. \cite{Palmer_2004}, Palmer and Stavrinou introduced the non-integer spherical Laplacian as the one in our Eq. (\ref{eq3.8}) above, but they stated that this form was obtained by assigning the non-integer dimension to $\alpha _{3}$, while it is in fact assigned to $\alpha _{2}$. In our paper I, we used this same form of the spherical Laplacian to discuss the fractional-dimension solutions to the Laplace equation and the related multipole expansion (see Appendix A of Ref. \cite{Varieschi:2020ioh}), but we could have used also the other forms of the Laplacian discussed in this section. However, our main NFDG results in Eqs. (\ref{eq2.3})-(\ref{eq2.4}) are independent of the choice of the fractional-dimension Laplace operator.

From the general Laplacian in Eq. (\ref{eq3.6}), other ``mixed'' forms of this operator are possible. For example, the non-integer dimension could be equally distributed over the three parameters by setting $\alpha _{1} =\alpha _{2} =\alpha _{3} =D/3$, or in an unequal way, or over just two parameters, etc. Therefore, there is a certain ambiguity in how the non-integer dimension is acting over the three spatial coordinates, as we already remarked in Sect. \ref{sect:Mathematical} above. We also note that the order of the parameters, $\alpha _{1}$, $\alpha _{2}$, $\alpha _{3}$, refers to the original weights in Eq. (\ref{eq2.2}), which were related to rectangular coordinates and not to the spherical coordinates used in this section.

\subsection{\label{subsect:cylindrical}Cylindrical coordinates
}
In cylindrical coordinates $R$, $\varphi $, $z$, we can use the orthonormal basis $ \partial _{\mu } =\left (\frac{ \partial }{ \partial t} ,\frac{ \partial }{ \partial R} ,\frac{1}{R}\frac{ \partial }{ \partial \varphi } ,\frac{ \partial }{ \partial z}\right )$ and the same Lagrangian density $\mathcal{L} = \nabla \phi ^{ \ast } \cdot  \nabla \phi  = \partial _{i}\phi ^{ \ast } \partial _{i}\phi $ in the main Eq. (\ref{eq3.2}). This time, we obtain:

\begin{gather} \nabla _{\alpha _{1} ,\alpha _{2} ,\alpha _{3}}^{2}\phi \left (R ,\varphi  ,z\right ) =\left [\frac{ \partial ^{2}\phi }{ \partial R^{2}} +\frac{\left (\alpha _{1} +\alpha _{2} -1\right )}{R}\frac{ \partial \phi }{ \partial R}\right ] \label{eq3.10} \\
 +\frac{1}{R^{2}}\left [\frac{ \partial ^{2}\phi }{ \partial \varphi ^{2}} +\frac{\left (\alpha _{2} -1\right )}{\tan \varphi }\frac{ \partial \phi }{ \partial \varphi } +\frac{\left (1 -\alpha _{1}\right )}{\cot \varphi }\frac{ \partial \phi }{ \partial \varphi }\right ] +\left [\frac{ \partial ^{2}\phi }{ \partial z^{2}} +\frac{\left (\alpha _{3} -1\right )}{\ensuremath{\operatorname*{}}z}\frac{ \partial \phi }{ \partial z}\right ] . \nonumber \end{gather}This is the most general cylindrical Laplacian for $D =\alpha _{1} +\alpha _{2} +\alpha _{3}$ ($0 <\alpha _{1} ,\alpha _{2} ,\alpha _{3} \leq 1$). For $\alpha _{1} =\alpha _{2} =\alpha _{3} =1$ ($D =3$), the standard cylindrical Laplacian is recovered, while special cases are obtained if the non-integer dimension is assigned to just one of the three parameters, as for the spherical case studied in the previous subsection.

If the non-integer parameter is the first one, that is $0 <\alpha _{1} <1$, $\alpha _{2} =\alpha _{3} =1$, $D =\alpha _{1} +2$, we have:
\begin{gather} \nabla _{D -2 ,1 ,1}^{2}\phi \left (R ,\varphi  ,z\right ) =\left [\frac{ \partial ^{2}\phi }{ \partial R^{2}} +\frac{\left (D -2\right )}{R}\frac{ \partial \phi }{ \partial R}\right ] \label{eq3.11} \\
 +\frac{1}{R^{2}}\left [\frac{ \partial ^{2}\phi }{ \partial \varphi ^{2}} +\frac{\left (3 -D\right )}{\cot \varphi }\frac{ \partial \phi }{ \partial \varphi }\right ] +\left [\frac{ \partial ^{2}\phi }{ \partial z^{2}}\right ] . \nonumber \end{gather}If instead, $0 <\alpha _{2} <1$, $\alpha _{1} =\alpha _{3} =1$, $D =\alpha _{2} +2$, we have:
\begin{gather} \nabla _{1 ,D -2 ,1}^{2}\phi \left (R ,\varphi  ,z\right ) =\left [\frac{ \partial ^{2}\phi }{ \partial R^{2}} +\frac{\left (D -2\right )}{R}\frac{ \partial \phi }{ \partial R}\right ] \label{eq3.12} \\
 +\frac{1}{R^{2}}\left [\frac{ \partial ^{2}\phi }{ \partial \varphi ^{2}} +\frac{\left (D -3\right )}{\tan \varphi }\frac{ \partial \phi }{ \partial \varphi }\right ] +\left [\frac{ \partial ^{2}\phi }{ \partial z^{2}}\right ] . \nonumber \end{gather}Finally, if $0 <\alpha _{3} <1$, $\alpha _{1} =\alpha _{2} =1$, $D =\alpha _{3} +2$,  we obtain:
\begin{gather} \nabla _{1 ,1 ,D -2}^{2}\phi \left (R ,\varphi  ,z\right ) =\left [\frac{ \partial ^{2}\phi }{ \partial R^{2}} +\frac{1}{R}\frac{ \partial \phi }{ \partial R}\right ] \label{eq3.13} \\
 +\frac{1}{R^{2}}\left [\frac{ \partial ^{2}\phi }{ \partial \varphi ^{2}}\right ] +\left [\frac{ \partial ^{2}\phi }{ \partial z^{2}} +\frac{D -3}{z}\frac{ \partial \phi }{ \partial z}\right ] . \nonumber \end{gather} 

From the general cylindrical Laplacian in Eq. (\ref{eq3.10}), other ``mixed'' forms of this operator are possible. Again, the non-integer dimension could be equally distributed over the three parameters by setting $\alpha _{1} =\alpha _{2} =\alpha _{3} =D/3$, or in an unequal way, or over just two parameters, etc. In this cylindrical case, it is obvious that $\alpha _{3}$ refers directly to the $z$ coordinate, while it is not possible to assign $\alpha _{1}$ and $\alpha _{2}$ to the $R$, $\varphi $ coordinates. Therefore, a certain ambiguity remains in how to distribute the non-integer dimension over the three spatial coordinates also in this case.

\section{\label{sect::Relativistic} Relativistic equations for spaces with non-integer dimension
}
In Sect. \ref{sect::Euler}, it was shown that  the Euler-Lagrange equations for spaces with non-integer dimensions can be obtained by substituting $\mathcal{L} \rightarrow \prod \limits _{i =1}^{3}W_{1}\left (x^{i} ,\alpha _{i}\right )\mathcal{L}$, i.e., simply by multiplying the Lagrangian density by the product of the weights for the three spatial coordinates. This immediately suggests a possible procedure for the relativistic extension of NFDG: include the same weight factor $\prod \limits _{i =1}^{3}W_{1}\left (x^{i} ,\alpha _{i}\right )$ inside the standard Hilbert action $S_{H} =\int \sqrt{ -g}\ R\ d^{4}x$ and then vary this modified action with respect to the inverse metric $g^{\mu \nu }$, as it is usually done in standard GR.

This procedure is practically equivalent to the one used for scalar-tensor theories of gravity (see Ref. \cite{Carroll:2004st} for a general overview) and it has been used extensively by Calcagni in the context of multi-scale spacetimes and fractional gravity theories \cite{Calcagni:2011sz,Calcagni:2009kc,Calcagni:2010bj,Calcagni:2013yqa,Calcagni:2020ads,Calcagni:2021ipd,Calcagni:2021aap}. In the following subsections we will review these techniques and adapt them to our particular case.

\subsection{\label{subsect:relativisticone}RFDG field equations
}
In this section, we will obtain the field equations by following closely the methods used by Calcagni in his main paper on multi-scale gravity and cosmology \cite{Calcagni:2013yqa} and the general procedure for field equations in alternative theories of gravity (see Sect. 4.8 in Ref. \cite{Carroll:2004st}). Following \cite{Calcagni:2010bj,Calcagni:2013yqa}, the weight factor $\prod \limits _{i =1}^{3}W_{1}\left (x^{i} ,\alpha _{i}\right )$ introduced in Sect. \ref{sect::Euler}, with the NFDG weights from Eq. (\ref{eq2.2}), is consistent with the general form of the weight $v\left (x\right )$, assumed to be factorizable in the coordinates and positive semi-definite \cite{Calcagni:2013yqa}:

\begin{gather}v\left (x\right ) =\prod \limits _{\mu  =0}^{3}v_{\mu }\left (x^{\mu }\right ) ,\ v_{\mu }\left (x^{\mu }\right ) \geq 0 \label{eq4.1} \\
q^{\mu }\left (x^{\mu }\right ) =\int ^{x^{\mu }}dx^{ \prime ^{\mu }}v_{\mu }\left (x^{ \prime ^{\mu }}\right ) \nonumber \end{gather}
as shown in the first line of the previous equation.\protect\footnote{
We prefer to indicate explicitly the spacetime dimension (i.e., $D_{spacetime} =4$, $\mu  =0 ,1 ,2 ,3$), as opposed to using the symbol $D$ as in Ref. \cite{Calcagni:2013yqa}. We will continue instead to denote with $D$ the variable NFDG space dimension, as was done in Sects. \ref{sect:intro}-\ref{sect::Euler}.
}

The action measure is assumed to be of the form $d\varrho \left (x\right ) =d^{4}x\ \ v\left (x\right ) =d^{4}q\left (x\right )$, where ``geometric coordinates'' $q\left (x\right )$, as defined in the second line of Eq. (\ref{eq4.1}), can be used formally to re-express the measure in a standard Lebesgue form. In this way \cite{Calcagni:2013yqa}, a multi-scale Minkowski spacetime is defined as the multiplet $\mathcal{M}^{4} =\left (M^{4} ,\varrho  , \partial  ,\mathcal{K}\right )$ based on an ordinary $4$-dimensional Minkowski spacetime $M^{4}$, a Lebesgue-Stieltjes measure $\varrho $ for the action, a set of calculus rules with derivative operators $ \partial $, and an appropriate Laplace-Beltrami operator $\mathcal{K}$.

Different multi-scale theories were then developed by Calcagni, with reference to the possible derivative operators $ \partial $ being used: theory $T_{1}$ with ordinary derivatives, theory $T_{v}$ with weighted derivatives, and theory $T_{q}$ with q-derivatives (see \cite{Calcagni:2013yqa,Calcagni:2021ipd} for full details). These models were then used in connection with the most general measure derived from first principles \cite{Calcagni:2016xtk} and then applied to quantum field theories, quantum gravity, and cosmology \cite{Calcagni:2013yqa,Calcagni:2020ads,Calcagni:2021ljs,Calcagni:2021ipd,Calcagni:2021aap}.

For the purpose of deriving the RFDG field equations, we will consider the NFDG weight $v\left (x\right )$:

\begin{equation}v\left (x\right ) =\prod \limits _{\mu  =0}^{3}v_{\mu }\left (x^{\mu }\right ) =\prod \limits _{i =1}^{3}\frac{\pi ^{\alpha _{i}/2}}{\Gamma \left (\alpha _{i}/2\right )}\left \vert x^{i}\right \vert ^{\alpha _{i} -1} , \label{eq4.2}
\end{equation}consistent with Eqs. (\ref{eq2.2}) and (\ref{eq3.1}) and with the time weight assumed to be unity, i.e., $v_{0}\left (x^{0}\right ) =1$, but more general expressions can be used, including non-trivial time weights. As already noted in Appendix A of our paper III, using rescaled coordinates $w^{i} =x^{i}/l_{0}$, the NFDG\ weight $\frac{\pi ^{\alpha _{i}/2}}{\Gamma \left (\alpha _{i}/2\right )}\genfrac{\vert }{\vert }{}{}{x^{i}}{l_{0}}^{\alpha _{i} -1}$  in Eq. (\ref{eq4.2}) is very similar to the binomial weight $\left (1 +\genfrac{\vert }{\vert }{}{}{x^{i}}{l_{ \ast }}^{\alpha _{i} -1}\right )$ used by Calcagni \cite{Calcagni:2016xtk}. However, in NFDG the transition from Newtonian to non-Newtonian behavior is achieved by varying continuously the fractional dimension parameters $\alpha _{i}$ from $\alpha _{i} =1$ (Newtonian case, $v_{i}\left (x^{i}\right ) =1$) to $0 <\alpha _{i} <1$ (non-Newtonian), with $l_{0}$ being an appropriate scale parameter linked to the MOND acceleration scale. In multifractional theories, the scale lengths $l_{ \ast }$ represent the observation scales at which the spacetime dimension may change, with different behaviors for $x^{i} \ll l_{ \ast }$ and $x^{i} \gg l_{ \ast }$,  and with a binomial weight which does not simply reduce to unity for $\alpha _{i} =1$.

Apart from the different choice of weights, the RFDG field equations are obtained with the same procedure for the theory $T_{1}$ with ordinary derivatives \cite{Calcagni:2013yqa}. The action for gravity can be taken as:

\begin{equation}S_{g} =\frac{1}{16\pi G}\int d^{4}x\ \sqrt{ -g}v\left (x\right )\left [R -\omega  \partial _{\mu }v \partial ^{\mu }v -U\left (v\right )\right ] , \label{eq4.3}
\end{equation}
where $G$ is Newton's gravitational constant, $v\left (x\right )$ is the weight being considered, $g =\left \vert g_{\mu \nu }\right \vert $ is the determinant of the metric, $R =R_{\ \mu }^{\mu } =g^{\mu \nu }R_{\mu \nu }$ is the Ricci scalar, defined in terms of standard Ricci and Riemann tensors \cite{Carroll:2004st}. In scalar-tensor and multifractional theories, it is customary to include in the gravitational action a ``kinetic'' term $\omega  \partial _{\mu }v \partial ^{\mu }v$ and a ``potential'' term $U\left (v\right )$ (which can be set to $2\Lambda $, to include a cosmological constant $\Lambda $, or can be a function of the weight $v$). In general, these terms are not needed in RFDG, and we will set $\omega  =0$ and $U\left (v\right ) =0$ later.

Including also a matter action $S_{m} =\int d^{4}x\sqrt{ -g}v\left (x\right )\mathcal{L}_{m}$, with $\mathcal{L}_{m}$ denoting an appropriate Lagrangian density, the energy-momentum tensor is now defined as:

\begin{equation}T_{\mu \nu } = -\frac{2}{\sqrt{ -g}\ v\left (x\right )}\frac{\delta S_{m}}{\delta g^{\mu \nu }} , \label{eq4.4}
\end{equation}with the weight $v\left (x\right )$ added at the denominator. One can then obtain the field equations by varying the action with respect to the inverse metric $g^{\mu \nu }$, where additional terms are derived by using the techniques used for scalar-tensor models \cite{Carroll:2004st}. The final result is \cite{Calcagni:2013yqa}:

\begin{equation}R_{\mu \nu } -\frac{1}{2}g_{\mu \nu }\left [R -U\left (v\right )\right ] +g_{\mu \nu }\frac{ \square v}{v} -\frac{ \nabla _{\mu } \nabla _{\nu }v}{v} +\omega \left (\frac{1}{2}g_{\mu \nu } \partial _{\sigma }v \partial ^{\sigma }v - \partial _{\mu }v \partial _{\nu }v\right ) =8\pi GT_{\mu \nu } \label{eq4.5}
\end{equation}
where $ \nabla _{\mu }$ indicates standard covariant differentiation, and the Laplace-Beltrami operator is defined as $ \square  = \nabla ^{\mu } \nabla _{\mu } =g^{\mu \nu } \nabla _{\mu } \nabla _{\nu }$. It is easy to check that standard GR\ field equations are recovered for $\omega  =0$ and $v\left (x\right ) =1$, including a cosmological constant term by setting $U =2\Lambda $, or otherwise by simply setting $U =0$.

The trace of Eq. (\ref{eq4.5}) yields:

\begin{equation} -R +2U\left (v\right ) +3\frac{ \square v}{v} +\omega  \partial _{\mu }v \partial ^{\mu }v =8\pi GT_{\mu }^{\ \mu } \label{eq4.6}
\end{equation}while variation of the total action $S =S_{g} +S_{m}$ with respect to the weight $v\left (x\right )$ gives:

\begin{equation}R -U\left (v\right ) = -16\pi G\mathcal{L}_{m} +v\frac{dU}{dv} -\omega \left (2v \square v + \partial _{\mu }v \partial ^{\mu }v\right ) . \label{eq4.7}
\end{equation}
Combining the last two equations, one can also obtain \cite{Calcagni:2013yqa}:

\begin{equation}R -2v\frac{dU}{dv} +3\frac{ \square v}{v} +\omega \left (4v \square v +3 \partial _{\mu }v \partial ^{\mu }v\right ) =8\pi G\left (T_{\mu }^{\ \mu } -4\mathcal{L}_{m}\right ) \label{eq4.8}
\end{equation}
which links directly the Ricci scalar $R$ with the weight $v\left (x\right )$.

An alternative version of the field equation (\ref{eq4.5}), can be obtained by taking the trace of this equation and then combining the result with the same Eq. (\ref{eq4.5}). The final result is:

\begin{equation}R_{\mu \nu } =8\pi G\left (T_{\mu \nu } -\frac{1}{2}g_{\mu \nu }T\right ) +\frac{1}{2}g_{\mu \nu }U\left (v\right ) +\frac{1}{2}g_{\mu \nu }\frac{ \square v}{v} +\frac{ \nabla _{\mu } \nabla _{\nu }v}{v} +\omega \left . \partial _{\mu }v \partial _{\nu }v\right ., \label{eq4.9}
\end{equation}where $T =T_{\ \mu }^{\mu }$ is the trace of the energy-momentum tensor.

The RFDG\ equations can be obtained from the previous general equations (\ref{eq4.3})-(\ref{eq4.9}) by setting $\omega  =0$, $U\left (v\right ) =0$, and by using the NFDG weight $v\left (x\right )$ described in Eq. (\ref{eq4.2}), or any other appropriate weight. In particular, the field equation becomes:
\begin{equation}R_{\mu \nu } -\frac{1}{2}g_{\mu \nu }\left .R\right . +g_{\mu \nu }\frac{ \square v}{v} -\frac{ \nabla _{\mu } \nabla _{\nu }v}{v} =8\pi GT_{\mu \nu } \label{eq4.10}
\end{equation}
where only the two additional terms $g_{\mu \nu }\frac{ \square v}{v} -\frac{ \nabla _{\mu } \nabla _{\nu }v}{v}$ in the left-hand side of this equation need to be computed, in order to extend standard GR to RFDG. The alternative version, corresponding to Eq. (\ref{eq4.9}), is instead:
\begin{equation}R_{\mu \nu } =8\pi G\left (T_{\mu \nu } -\frac{1}{2}g_{\mu \nu }T\right ) +\frac{1}{2}g_{\mu \nu }\frac{ \square v}{v} +\frac{ \nabla _{\mu } \nabla _{\nu }v}{v} , \label{eq4.11}
\end{equation}which will be used in the next section to derive the Friedmann equations of cosmology.

Following the discussion in Sect. 3.1 of Ref. \cite{Calcagni:2013yqa}, we note that the weight $v\left (x\right )$ can be treated as a scalar field in the derivation of the field equations \cite{Carroll:2004st,CLIFTON20121}, but it should be considered a ``fixed coordinate profile'' and not a Lorentz scalar. The derivation of the field equations is essentially equivalent to the one typically used in scalar-tensor theories \cite{Carroll:2004st,CLIFTON20121}, but the interpretation \cite{Calcagni:2013yqa} of the scalar weight $v\left (x\right )$ differs from the one of the fields $\phi\left (x\right )$ used in modified gravity and in quintessence models of dark energy \cite{Tsujikawa:2013fta}.

Although $v\left (x\right )$ does not represent a dynamical field, it affects the dynamics through the additional terms $g_{\mu \nu }\frac{ \square v}{v} -\frac{ \nabla _{\mu } \nabla _{\nu }v}{v}$ in Eq. (\ref{eq4.10}) above. Since our weight $v\left (x\right )$  in Eq. (\ref{eq4.2}) is determined directly by our NFDG theory, we do not feel necessary, at least at this stage, to introduce kinetic and potential terms, $\omega  \partial _{\mu }v \partial ^{\mu }v$ and $U\left (v\right )$, as was done in multifractional gravitational theories \cite{Calcagni:2013yqa}.

Therefore, at least at this stage, RFDG is introduced in a phenomenological way by fixing from the beginning the coordinate profile or weight $v\left (x\right )$, which does not change while the system is evolving dynamically. The choice of the weight is suggested by those used in our previous NFDG papers, or by similar time-dependent weights which will be used in the next sub-section. As already mentioned above, $v\left (x\right )$ cannot be considered a scalar field, although the derivation of the field equations is equivalent to the one for scalar-tensor theories (see also Sect 3.1 in Ref. \cite{CLIFTON20121}). The RFDG field equations (\ref{eq4.10}) and (\ref{eq4.11}) obviously reduce to standard GR for $v\left (x\right )=1$, i.e., for gravitational systems which do not possess any spatial or temporal fractional-dimension (for example, at the Solar System level). Therefore, RFDG and GR are fully consistent for structures whose Hausdorff dimension coincides with the topological one. 

In the next subsection, we will apply the main field equations (\ref{eq4.10}) and (\ref{eq4.11}) to the case of standard cosmology and to the Friedmann-Lema{\^\i}tre-Robertson-Walker metric.

\subsection{\label{subsect:relativistictwo}  Cosmology and RFDG
}
 In standard cosmology \cite{Carroll:2004st,Carroll:2000fy}, the Friedmann-Lema{\^\i}tre-Robertson-Walker (FLRW) metric is usually expressed as ($c =1$):

\begin{equation}ds^{2} = -dt^{2} +a^{2}\left (t\right )\left [\frac{dr^{2}}{1 -\kappa r^{2}} +r^{2}d\Omega ^{2}\right ] , \label{eq4.12}
\end{equation}where $a\left (t\right ) =R\left (t\right )/R_{0}$ is the dimensionless scale factor ($R\left (t\right )$ is the scale factor, $R_{0} =R\left (t_{0}\right )$, $t_{0}$ current time), $\kappa  =k/R_{0}^{2}$ ($k = -1$ open universe; $k =0$ flat universe; $k =1$ closed universe), and $d\Omega ^{2} =d\theta ^{2} +\sin ^{2}\theta d\varphi ^{2}$. Following this choice for the FLRW metric, the Christoffel symbols, the non-zero components of the Ricci tensor, and the Ricci scalar are readily computed \cite{Carroll:2004st} and are reported in Appendix \ref{sect::appendix_one}.

Matter and energy in the Universe are usually modeled as a perfect fluid with energy-momentum tensor:

\begin{equation}T_{\mu \nu } =\left (\rho  +p\right )U_{\mu }U_{\nu } +pg_{\mu \nu } \label{eq4.13}
\end{equation}
with the fluid at rest in comoving coordinates, so that the four-velocity is $U^{\mu } =\left (1 ,0 ,0 ,0\right )$ and the energy-momentum tensor simply becomes

\begin{equation}T_{\mu \nu } =\left (\begin{array}{cccc}\rho  & 0 & 0 & 0 \\
0 & \, & \, & \, \\
0 & \, & g_{ij}p & \, \\
0 & \, & \, & \,\end{array}\right ) \label{eq4.14}
\end{equation}
in terms of the energy density $\rho \left (t\right )$ and the pressure $p\left (t\right )$. This can also be written as $T_{\ \nu }^{\mu } =diag\left ( -\rho  ,p ,p ,p\right )$  and with the trace given by $T =T_{\ \mu }^{\mu } = -\rho  +3p$.

In order to compute the additional terms $g_{\mu \nu }\frac{ \square v}{v}$, $\frac{ \nabla _{\mu } \nabla _{\nu }v}{v}$ in Eqs. (\ref{eq4.10})-(\ref{eq4.11}), we should express the NFDG weight of Eq. (\ref{eq4.2}) in terms of spherical coordinates $r$, $\theta $, $\varphi $. Using standard coordinate transformations between rectangular and spherical coordinates and assuming for example $\alpha _{1} =\alpha _{2} =\alpha _{3} =D/3$, we obtain:
\begin{equation}v\left (x\right ) =\prod \limits _{i =1}^{3}\frac{\pi ^{\alpha _{i}/2}}{\Gamma \left (\alpha _{i}/2\right )}\left \vert x^{i}\right \vert ^{\alpha _{i} -1} =\frac{\pi ^{D/2}}{\left [\Gamma \left (D/6\right )\right ]^{3}}r^{D -3}\left \vert \sin \theta \right \vert ^{\left (\frac{2}{3}D -2\right )}\left \vert \cos \theta \right \vert ^{\left (\frac{D}{3} -1\right )}\left \vert \sin \varphi \right \vert ^{\left (\frac{D}{3} -1\right )}\left \vert \cos \varphi \right \vert ^{\left (\frac{D}{3} -1\right )} =v_{r}\left (r\right )v_{\theta }\left (\theta \right )v_{\varphi }\left (\varphi \right ) , \label{eq4.15}
\end{equation}but this weight does not yield isotropic results for the Friedmann equations. Assuming instead a simpler radial weight $v_{r}\left (r\right ) =\left .\frac{\pi ^{\left (D/2 -1\right )}}{2\Gamma \left (D/2\right )}r^{D -3}\right .$, which follows from the general fractional-dimension integral in Eq. (\ref{eq2.1}), divided by the standard factor of $4\pi r^{2}$ pertaining to the $D =3$ case, still does not seem to yield isotropic results due to the presence of mixed $\left (t ,r\right )$ components in the field tensors, which can be avoided only by adding a time weight $v_{t}\left (t\right ) =a\left (t\right )$, equal to the scale factor.\protect\footnote{
Even using a combined weight, $v_{t}\left (t\right )v_{r}\left (r\right ) =a\left (t\right )v_{r}\left (r\right )$, does not seem to yield fully isotropic Friedmann equations for the cosmological problem. A more detailed study of cosmological weights, including possible radial factors or even direct modifications to the FLRW metric in terms of variable space-time dimensions, will be done in a future publication.
}

As discussed in Appendix \ref{sect::appendix_one}, it might be more appropriate for cosmological applications to assume a purely temporal weight, similar to the spatial one in Eq. (\ref{eq4.2}):
\begin{equation}v\left (x\right ) \equiv v_{t}\left (t\right ) =\frac{\pi ^{\alpha _{t}/2}}{\Gamma \left (\alpha _{t}/2\right )}t^{\alpha _{t} -1} , \label{eq4.16}
\end{equation}where $t >0$ and $0 <\alpha _{t} \leq 1$ is a time fractional dimension. This assumption is similar to the one used by Calcagni in his Ref. \cite{Calcagni:2010bj}, but will yield different results in the context of RFDG.

With the particular weight in Eq. (\ref{eq4.16}), all quantities in the generalized field equations (\ref{eq4.5}) and (\ref{eq4.9}) can be computed and the complete results are detailed in Appendix \ref{sect::appendix_one}. Using these results, the modified Friedmann equations are:

\begin{gather}\genfrac{(}{)}{}{}{\overset{ .}{a}}{a}^{2} +\frac{\overset{ .}{a}\overset{ .}{v}}{av} -\frac{\omega \overset{ .}{v}^{2}}{6} -\frac{U\left (v\right )}{6} =\frac{8\pi G}{3}\rho  -\frac{\kappa }{a^{2}} \label{eq4.17} \\
\frac{\overset{ . .}{a}}{a} +\frac{\overset{ .}{a}\overset{ .}{v}}{2av} +\frac{\overset{ . .}{v}}{2v} +\frac{\omega \overset{ .}{v}^{2}}{3} -\frac{U\left (v\right )}{6} = -\frac{4\pi G}{3}\left (\rho  +3p\right ) \nonumber \end{gather}where we denoted the temporal weight simply as $v =v_{t}\left (t\right )$ and all time derivatives are shown using the over-dot notation.

These equations can be further simplified by taking $\omega  =0$ and by introducing a possible cosmological constant $\Lambda $ (setting $U\left (v\right ) =2\Lambda $), for comparison with standard $\Lambda CDM$ cosmology. Therefore, we obtain:
\begin{gather}\genfrac{(}{)}{}{}{\overset{ .}{a}}{a}^{2} +\frac{\overset{ .}{a}\overset{ .}{v}}{av} =\frac{8\pi G}{3}\rho  -\frac{\kappa }{a^{2}} +\frac{\Lambda }{3} \label{eq4.18} \\
\frac{\overset{ . .}{a}}{a} +\frac{\overset{ .}{a}\overset{ .}{v}}{2av} +\frac{\overset{ . .}{v}}{2v} = -\frac{4\pi G}{3}\left (\rho  +3p\right ) +\frac{\Lambda }{3} \nonumber \end{gather}which can be compared directly with the standard Friedmann equations \cite{Carroll:2004st,Carroll:2000fy}. It is evident that both Eqs. (\ref{eq4.17})-(\ref{eq4.18}) reduce to the standard ones for $v =1$ ($\overset{ .}{v} =\overset{ . .}{v} =0$).

The Hubble parameter $H$ characterizes the rate of expansion, as usual:

\begin{gather}H =\frac{\overset{ .}{a}}{a} \label{eq4.19} \\
\overset{ .}{H} =\frac{\overset{ . .}{a}}{a} -\genfrac{(}{)}{}{}{\overset{ .}{a}}{a}^{2} =\frac{\overset{ . .}{a}}{a} -H^{2} \nonumber \end{gather}
with the present epoch value as the Hubble constant $H_{0} =100h\ \mbox{km}\ \mbox{s}^{ -1}\ \mathrm{M}\mathrm{p}\mathrm{c}^{ -1}$ ($h \approx 0.7$). In a similar way, we can introduce a weight parameter $V$:
\begin{gather}V =\frac{\overset{ .}{v}}{v} \label{eq4.20} \\
\overset{ .}{V} =\frac{\overset{ . .}{v}}{v} -\genfrac{(}{)}{}{}{\overset{ .}{v}}{v}^{2} =\frac{\overset{ . .}{v}}{v} -V^{2} \nonumber \end{gather}
and rewrite the Friedmann equations (\ref{eq4.18}) in terms of the $H$ and $V$ parameters:
\begin{gather}H^{2} +HV =\frac{8\pi G}{3}\rho  -\frac{\kappa }{a^{2}} +\frac{\Lambda }{3} \label{eq4.21} \\
\overset{ .}{H} +H^{2} +\frac{1}{2}\left (HV +V^{2} +\overset{ .}{V}\right ) = -\frac{4\pi G}{3}\left (\rho  +3p\right ) +\frac{\Lambda }{3} . \nonumber \end{gather}

We will assume that the standard components of the Universe have energy densities evolving as power laws, $\rho _{i}\left (t\right ) =\rho _{i0}a^{ -n_{i}}\left (t\right )$; each component will have equation of state $p_{i}\left (t\right ) =w_{i}\rho _{i}\left (t\right )$, with parameters $w_{i} =\frac{1}{3}n_{i} -1$. As in standard cosmology, we will include matter ($M$, $n_{M} =3$, $w_{M} =0$), radiation ($R$, $n_{R} =4$, $w_{R} =\frac{1}{3}$), curvature ($C$, $n_{C} =2$, $w_{C} = -\frac{1}{3}$), and vacuum ($\Lambda $, $n_{\Lambda } =0$, $w_{\Lambda } = -1$) .

Generalizing the standard procedure used in $\Lambda CDM$ cosmology \cite{Carroll:2004st,Carroll:2000fy}, we will still introduce the density parameter $\Omega $ and the critical density $\rho _{crit}$ as $\Omega  =\frac{8\pi G}{3H^{2}}\rho  =\frac{\rho }{\rho _{crit}}$ and $\rho _{crit} =\frac{3H^{2}}{8\pi G}$, respectively. These two equations assume that for each component the density parameter is defined as $\Omega _{i} =\frac{8\pi G}{3H^{2}}\rho _{i} =\frac{\rho _{i}}{\rho _{crit}}$, with the special cases for the curvature energy density $\rho _{C} \equiv  -\frac{3\kappa }{8\pi Ga^{2}}$ and the vacuum energy density $\rho _{\Lambda } \equiv \frac{\Lambda }{8\pi G}$. While the curvature density parameter $\Omega _{C} = -\frac{\kappa }{H^{2}a^{2}}$ is typically not included in the total $\Omega  =\Omega _{M} +\Omega _{R} +\Omega _{\Lambda }$ introduced above, it is still possible to modify the first Friedmann equation (\ref{eq4.21}) and obtain:

\begin{gather}1 +\beta  =\Omega  -\frac{\kappa }{H^{2}a^{2}} =\Omega _{M} +\Omega _{R} +\Omega _{\Lambda } +\Omega _{C} =\sum _{i}\Omega _{i} \label{eq4.22} \\
\beta  \equiv \frac{V}{H} \nonumber \end{gather}
which extends the standard relation $\Omega  -1 =\frac{\kappa }{H^{2}a^{2}}$ and with the summation in the first line applied to all four components.\protect\footnote{
In RFDG, the connection with open ($\kappa  <0$), flat ($\kappa  =0$), and closed ($\kappa  >0$) universes is not simply related to the density parameter $\Omega  \lesseqqgtr 1$ as in standard cosmology, due to the presence of the additional $\beta $ term in Eq. (\ref{eq4.22}).
}

For the current time $t_{0}$, Eq. (\ref{eq4.22}) can be written as $\sum _{i}\Omega _{i0} =1 +\beta _{0}$, with $\beta _{0} =\frac{V_{0}}{H_{0}}$, and used to rewrite the first line in Eq. (\ref{eq4.21}) as:

\begin{gather}H^{2} +HV =\frac{8\pi G}{3}\sum _{i}\rho _{i}\left (t\right ) =\frac{8\pi G}{3}\sum _{i}\rho _{i0}a^{ -n_{i}}\left (t\right ) =H_{0}^{2}\sum _{i}\Omega _{i0}a^{ -n_{i}}\left (t\right ) \label{eq4.23} \\
 =H_{0}^{2}\{\left .\Omega _{M0}a^{ -3}\left (t\right ) +\Omega _{R0}a^{ -4}\left (t\right ) +\Omega _{\Lambda 0} +\left [1 +\beta _{0} -\left (\Omega _{M0} +\Omega _{R0} +\Omega _{\Lambda 0}\right )\right ]\right .a^{ -2}\left (t\right )\} , \nonumber \end{gather}where the summation symbols include all four components of the energy density, the current-time curvature density parameter is expressed in terms of the other three, $\Omega _{C0} =1 +\beta _{0} -\left (\Omega _{M0} +\Omega _{R0} +\Omega _{\Lambda 0}\right )$, and the explicit values of the integer parameters $n_{i}$ have also been used in the last line.

It is customary to use a dimensionless time $\overline{t} =H_{0}\left (t -t_{0}\right )$ when solving the previous differential equation, so we need to rewrite the RFDG weight in Eq. (\ref{eq4.16}) in terms of $t =t_{0} +\frac{\overline{t}}{H_{0}}$ and then rescale this variable for dimensional correctness, dividing by a scale time $t_{sc}$ which can be simply taken as the current time, i.e., $t_{sc} \approx t_{0}$. Then, we have:
\begin{gather}\frac{t}{t_{sc}} =\frac{t_{0}}{t_{sc}} +\frac{\overline{t}}{t_{sc}H_{0}} \approx 1 +\frac{\overline{t}}{t_{0}H_{0}} =1 +\delta _{0}\overline{t} \label{eq4.24} \\
v =v_{t}\left (\overline{t}\right ) =\frac{\pi ^{\alpha _{t}/2}}{\Gamma \left (\alpha _{t}/2\right )}\genfrac{(}{)}{}{}{t}{t_{sc}}^{\alpha _{t} -1} \approx \frac{\pi ^{\alpha _{t}/2}}{\Gamma \left (\alpha _{t}/2\right )}\left (1 +\delta _{0}\overline{t}\right )^{\alpha _{t} -1} \nonumber \end{gather}
and the final weight $v_{t}\left (\overline{t}\right )$ in the second line can be used with free parameters $\alpha _{t} >0$ and $\delta _{0} =\frac{1}{t_{0}H_{0}} \sim 1$, since typically $t_{0} \sim H_{0}^{ -1}$. With these approximations, we also find $\beta _{0} \approx \left (\alpha _{t} -1\right )$ and the only free parameter remaining in our equations is the time dimension $\alpha _{t}$.

Using the definitions for $H$ and $V$ from Eqs. (\ref{eq4.19})-(\ref{eq4.20}), the dimensionless time variable $\overline{t} =H_{0}\left (t -t_{0}\right )$ (with $d\overline{t} =H_{0}dt$), and with some additional algebra the main equation (\ref{eq4.23}) can be recast as:

\begin{equation}\overset{ .}{a} = -\frac{1}{2}\frac{a\overset{ .}{v}}{v} \pm a\sqrt{\bigg\{\left .\Omega _{M0}a^{ -3}\left (\overline{t}\right ) +\Omega _{R0}a^{ -4}\left (\overline{t}\right ) +\Omega _{\Lambda 0} +\left [1 +\beta _{0} -\left (\Omega _{M0} +\Omega _{R0} +\Omega _{\Lambda 0}\right )\right ]\right .a^{ -2}\left (\overline{t}\right )\bigg\} +\frac{1}{4}\genfrac{(}{)}{}{}{\overset{ .}{v}}{v}^{2}} , \label{eq4.25}
\end{equation}
which becomes the RFDG differential equation for the scale factor $a\left (\overline{t}\right )$ with the initial condition $a\left (0\right ) =1$ and time derivatives now taken with respect to $\overline{t}$. For an expanding universe at the current epoch, we will choose the positive sign in Eq. (\ref{eq4.25}), and then solve it numerically for any assumed values of $\Omega _{M0}$, $\Omega _{R0}$, $\Omega _{\Lambda 0}$ at the current time and for any given temporal weight $v =v_{t}\left (\overline{t}\right )$. It should be noted that for $\alpha _{t} =1$ and $\beta _{0} =0$   ($v =1$, $\overset{ .}{v} =0$, $\overset{ . .}{v} =0$), Eq. (\ref{eq4.25}) correctly reduces to the $\Lambda CDM$ equivalent differential equation:

\begin{equation}\overset{ .}{a} =a\sqrt{\{\left .\Omega _{M0}a^{ -3}\left (\overline{t}\right ) +\Omega _{R0}a^{ -4}\left (\overline{t}\right ) +\Omega _{\Lambda 0} +\left [1 -\left (\Omega _{M0} +\Omega _{R0} +\Omega _{\Lambda 0}\right )\right ]\right .a^{ -2}\left (\overline{t}\right )\}} , \label{eq4.26}
\end{equation}which is commonly used in standard cosmology to obtain $a\left (\overline{t}\right )$ from the initial $\Omega _{i0}$ values \cite{Carroll:2004st,Carroll:2000fy}.

\begin{figure}\centering 
\setlength\fboxrule{0in}\setlength\fboxsep{0.1in}\fcolorbox[HTML]{FFFFFF}{FFFFFF}{\includegraphics[ width=6.96in, height=4.598755555555555in,]{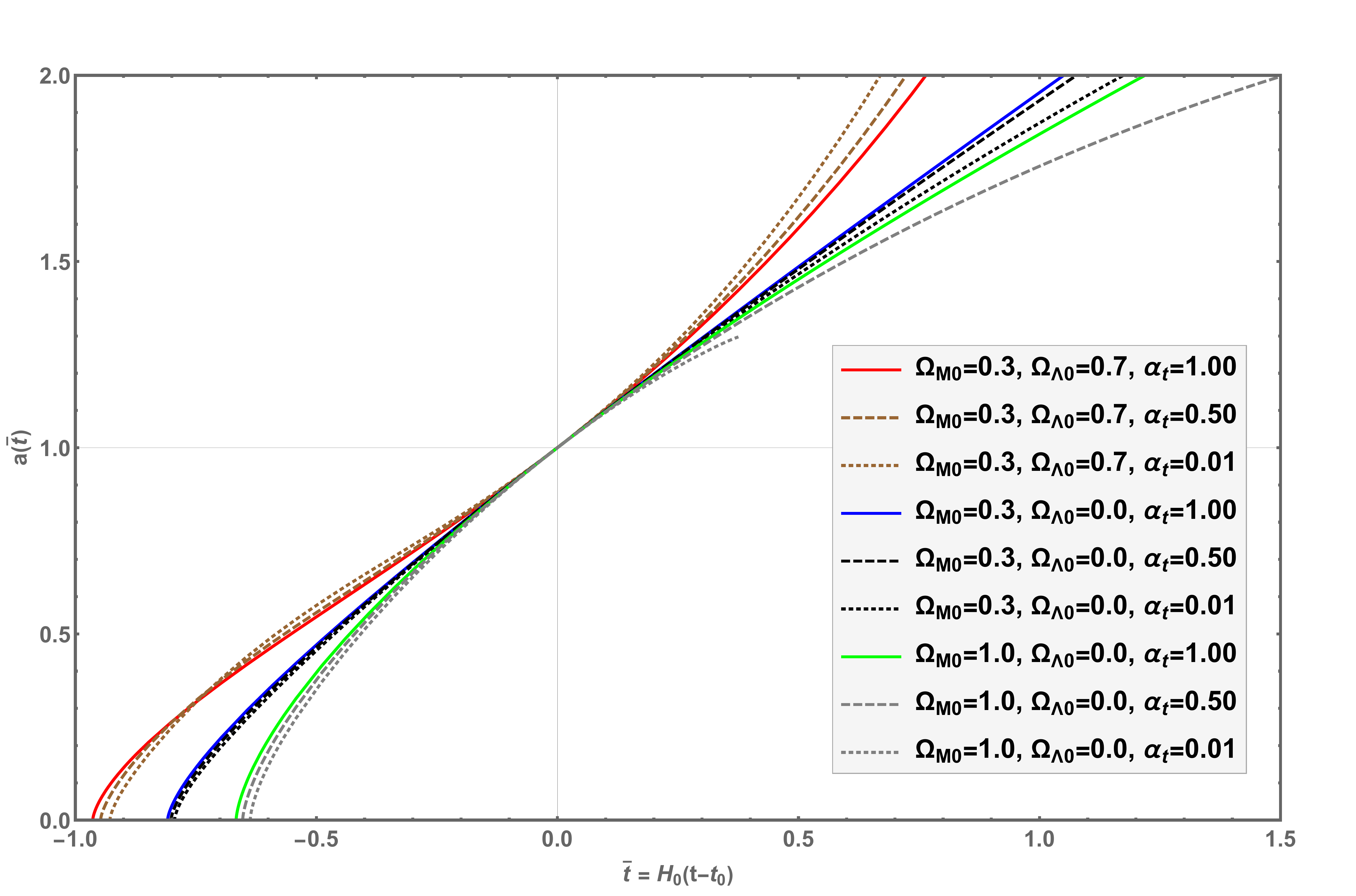}
}
\caption{Expansion histories for different values of $\Omega _{M0}$, $\Omega _{\Lambda 0}$, and of the RFDG parameter $\alpha _{t}$. Three notable cases from standard cosmology (red, blue, and green solid curves) are compared with RFDG results for similar $\Omega _{M0}$, $\Omega _{\Lambda 0}$ parameters, but with variable $\alpha _{t} >0$. RFDG curves for $\alpha _{t} =0.01$, $0.50$ (dotted and dashed curves) are only slightly different from their respective standard cosmology solid curves.}\label{figure:fig1}\end{figure}

Using the RFDG Friedmann equation (\ref{eq4.25}), or the standard-cosmology equivalent (\ref{eq4.26}) above, we plot  in Fig. \ref{figure:fig1} some results for different values of the parameters, using the temporal weight $v =v_{t}\left (\overline{t}\right )$ as described in Eq. (\ref{eq4.24}) with $0 <\alpha _{t} \leq 1$ and $\delta _{0} =\frac{1}{t_{0}H_{0}} \sim 1$. The results do not appear to depend much on the value of this second parameter, so we have simply set $\delta _{0} =1$\ \  in the following.

In this figure, we plot three notable standard cosmology expansion histories, similar to those presented in Fig. 8.3 of Ref. \cite{Carroll:2004st}, or Fig. 2 in Ref. \cite{Carroll:2000fy}. These were obtained by using Eq. (\ref{eq4.26}) above: the red-solid curve for $\Omega _{M0} =0.3$, $\Omega _{\Lambda 0} =0.7$ (and $\alpha _{t} =1$, i.e., $v =1$) represents the currently favored $\Lambda CDM$ expansion history for a universe dominated by about $70 \%$ of cosmological constant, Dark Energy (DE) component and only about $30 \%$ of matter component (baryonic and dark matter). The green-solid curve corresponds instead to a matter-dominated universe with $\Omega _{M0} =1.0$ and no cosmological constant, while the blue-solid curve corresponds to a $30 \%$ matter component, without any cosmological constant. The radiation component at current epoch is assumed to be negligible ($\Omega _{R0} \approx 0$), while the curvature component is fixed by $\Omega _{C0} =1 -\left (\Omega _{M0} +\Omega _{R0} +\Omega _{\Lambda 0}\right )$.

Using Eq. (\ref{eq4.25}), we also plotted RFDG expansion histories for the same values of the $\Omega _{M0}$, $\Omega _{\Lambda 0}$ parameters ($\Omega _{R0} =0$), but for different values of the parameter $\alpha _{t} =0.01$, $0.50$ (dotted and dashed curves). This was done to show how the RFDG curves, with $\alpha _{t} \approx 0 -1$, can modify the standard-cosmology histories by adding the temporal weight $v =v_{t}\left (\overline{t}\right )$ from Eq. (\ref{eq4.24}). The goal of our original NFDG \cite{Varieschi:2020ioh,Varieschi:2020dnd,2021MNRAS.503.1915V} was to show how the effect of adding a possible spatial fractional-dimension $D <3$ could replace the DM component in astrophysical structures. Therefore, the goal of RFDG should be to show that also the DE component in the Universe might be explained by a fractional-dimension effect, possibly related to the temporal dimension parameter $\alpha _{t} <1$.

However, as seen in the figure, the modified RFDG curves differ only slightly from the standard-cosmology curves, for the range of the $\alpha _{t}$ parameter being used. As a consequence, it seems unlikely that a RFDG curve with no cosmological constant ($\Omega _{\Lambda 0} =0$) and $0 <\alpha _{t} <1$ might be able to match the $\Lambda CDM$ red-solid curve, i.e., replacing DE with a fractional-dimension effect. Further analysis will be needed to check this possibility, by considering an extended range for the $\alpha _{t}$ parameter, using different approximations for $t_{SC}$ and $\delta _{0}$ in Eq. (\ref{eq4.24}), and possibly by also including the ``kinetic'' and ``potential'' terms in Eq. (\ref{eq4.17}) ($\omega  \neq 0$ and $U\left (v\right ) \neq 0$).

It is beyond the scope of this paper to expand these considerations any further, since the goal of this current work was just to introduce the main equations of Relativistic Fractional-Dimension Gravity, following the non-relativistic equations of our original NFDG. At the moment, RFDG is just a tentative modified gravity model which needs to be explored in more detail before it can be effectively applied to astrophysical objects or cosmological investigations. In the near future, we are planning to analyze measurements of the luminosity distance of type Ia supernovae with RFDG techniques, to see if our model can interpret these data without resorting to the DE component as in standard $\Lambda CDM$ cosmology. This would be a necessary condition for the viability of RFDG as an alternative model of gravity.

\section{\label{sect::conclusion}
Conclusion
}
In this work, we outlined a relativistic extension of our Newtonian Fractional-Dimension Gravity, which was developed to model the dynamics of galaxies without using any dark matter component. While the analysis of the NFDG model is still ongoing with additional galaxies being studied with these methods, it was important to show that NFDG admits a possible relativistic version, although at the moment it is not sure if this Relativistic Fractional-Dimension Gravity will be useful to address astrophysical or cosmological problems.

In this paper we have shown that a relativistic version can be derived from  the mathematical theory for spaces with non-integer dimensions, the extended Euler-Lagrange equations for scalar fields, and the existing  methods for scalar-tensor models of gravity, multi-scale spacetimes, and fractional gravity theories. The key element in all these methods is to include an appropriate coordinate weight in the spacetime metric used in both NFDG and RFDG. These weights will include the fractional-dimension parameters which characterize these theories and should be considered to be different from the scalar functions used in other models.

As a first, tentative application of RFDG, we applied it to the FLRW metric of standard cosmology, using a simple time-dependent weight. We have shown that it is straightforward to extend the standard Friedmann equations and to solve them numerically for different choices of the parameters. At this time, it is not possible to predict if these modified cosmological equations will be of any physical significance, in relation to the DE problem, or others.

Future work on the subject will
be needed to test this model against the cosmological paradigm, considering other possible weights which might be relevant in astrophysics and cosmology, and also including the cosmic late-time acceleration, distance indicators, type Ia supernovae data, etc., before RFDG can be considered a viable alternative theory of gravity.

\begin{acknowledgments}This work was supported by the Department of Physics, Loyola Marymount University, Los Angeles. The author wishes to acknowledge Dr. G. Calcagni for very useful advice regarding multifractional theories as well as other topics, and the anonymous reviewers for helpful comments and suggestions.

\end{acknowledgments}\appendix

\section{\label{sect::appendix_one}  RFDG tensors for the FLRW metric
}
In this section we will detail the RFDG tensors used for the study of the FLRW metric and related expansion histories discussed in Sect. \ref{subsect:relativistictwo}. All these tensor quantities were computed using Mathematica code.\protect\footnote{
Mathematica, Version 12.2.0.0, Wolfram Research Inc.
} These programs were tested by checking them against results for known cases (standard GR and others) and then extended to include the additional tensors described in Sect. \ref{subsect:relativisticone}.

The FLRW metric was defined in Eq. (\ref{eq4.12}), in terms of the dimensionless scale factor $a\left (t\right )$ and using standard spherical coordinates ($r$, $\theta $, $\varphi $); the energy-momentum tensor in Eqs. (\ref{eq4.13})-(\ref{eq4.14}), where pressure $p\left (t\right )$ and energy density $\rho \left (t\right )$ depend on the synchronous time $t$. The only additional input is the factorizable weight $v\left (x\right ) \equiv v_{t}\left (t\right )v_{r}\left (r\right )v_{\theta }\left (\theta \right )v_{\varphi }\left (\varphi \right )$, which in general can be a function of the four spacetime coordinates.

As already mentioned in Sect. \ref{subsect:relativistictwo}, this general form of the weight does not seem to yield isotropic Friedmann equations, and even considering simplified weights, such as $v\left (x\right ) \equiv v_{t}\left (t\right )v_{r}\left (r\right )$ or $v\left (x\right ) \equiv v_{r}\left (r\right )$ does not seem to yield the required symmetry, although future studies might be needed to explore these weights in more detail. Therefore, we opted to use a purely time dependent weight, $v\left (x\right ) \equiv v_{t}\left (t\right )$ and we computed all the tensors in terms of this general form for the weight, obtaining the following results.

Christoffel symbols, non-zero components of the Ricci tensor, and Ricci scalar (same as standard GR results \cite{Carroll:2004st}):

\begin{equation}\begin{array}{ll}\Gamma _{11}^{0} =\frac{a\overset{ .}{a}}{1 -\kappa r^{2}} & \Gamma _{11}^{1} =\frac{\kappa r}{1 -\kappa r^{2}} \\
\Gamma _{22}^{0} =a\overset{ .}{a}r^{2} & \Gamma _{33}^{0} =a\overset{ .}{a}r^{2}\sin ^{2}\theta  \\
\Gamma _{01}^{1} =\Gamma _{02}^{2} =\frac{\overset{ .}{a}}{a} & \Gamma _{03}^{3} =\frac{\overset{ .}{a}}{a} \\
\Gamma _{22}^{1} = -r\left (1 -\kappa r^{2}\right ) & \Gamma _{33}^{1} = -r\left (1 -\kappa r^{2}\right )\sin ^{2}\theta  \\
\Gamma _{12}^{2} =\frac{1}{r} & \Gamma _{13}^{3} =\frac{1}{r} \\
\Gamma _{33}^{2} = -\sin \theta \cos \theta  & \Gamma _{23}^{3} =\cot \theta \end{array} \label{eq6.1}
\end{equation}
\begin{gather}R_{00} = -3\frac{\overset{ . .}{a}}{a} \label{eq6.2} \\
R_{11} =\frac{a\overset{ . .}{a} +2\overset{ .}{a}^{2} +2\kappa }{1 -\kappa r^{2}} \nonumber  \\
R_{22} =r^{2}\left (a\overset{ . .}{a} +2\overset{ .}{a}^{2} +2\kappa \right ) \nonumber  \\
R_{33} =r^{2}\left (a\overset{ . .}{a} +2\overset{ .}{a}^{2} +2\kappa \right )\sin ^{2}\theta  \nonumber  \\
R =6\left [\frac{\overset{ . .}{a}}{a} +\genfrac{(}{)}{}{}{\overset{ .}{a}}{a}^{2} +\frac{\kappa }{a^{2}}\right ] \nonumber \end{gather}where time derivatives are indicated by the over-dot notation.

The additional tensors in Eqs. (\ref{eq4.5}) and (\ref{eq4.9}) are computed as follows. The potential term $A_{\mu \nu } \equiv \frac{1}{2}g_{\mu \nu }U\left (v\right )$ is easily computed from the metric components:

\begin{gather}A_{00} = -\frac{1}{2}U\left (v\right ) \label{eq6.3} \\
A_{11} =\frac{1}{2}\frac{a^{2}U\left (v\right )}{\left (1 -\kappa r^{2}\right )} \nonumber  \\
A_{22} =\frac{1}{2}r^{2}a^{2}U\left (v\right ) \nonumber  \\
A_{33} =\frac{1}{2}r^{2}a^{2}U\left (v\right )\sin ^{2}\theta  . \nonumber \end{gather}

The components of the tensor $B_{\mu \nu } \equiv g_{\mu \nu }\frac{ \square v}{v}$, calculated using the Laplace-Beltrami operator $ \square  = \nabla ^{\mu } \nabla _{\mu } =g^{\mu \nu } \nabla _{\mu } \nabla _{\nu }$, are as follows:

\begin{gather}B_{00} =\frac{3\overset{ .}{a}\overset{ .}{v} +a\overset{ . .}{v}}{av} \label{eq6.4} \\
B_{11} = -\frac{a\left (3\overset{ .}{a}\overset{ .}{v} +a\overset{ . .}{v}\right )}{v\left (1 -\kappa r^{2}\right )} \nonumber  \\
B_{22} = -\frac{r^{2}a\left (3\overset{ .}{a}\overset{ .}{v} +a\overset{ . .}{v}\right )}{v} \nonumber  \\
B_{33} = -\frac{r^{2}a\left (3\overset{ .}{a}\overset{ .}{v} +a\overset{ . .}{v}\right )\sin ^{2}\theta }{v} \nonumber \end{gather}
where the weight $v_{t}\left (t\right )$ is simply denoted by $v$. The tensor $C_{\mu \nu } \equiv \frac{ \nabla _{\mu } \nabla _{\nu }v}{v}$ has components:

\begin{gather}C_{00} =\frac{\overset{ . .}{v}}{v} \label{eq6.5} \\
C_{11} = -\frac{a\overset{ .}{a}\overset{ .}{v}}{v\left (1 -\kappa r^{2}\right )} \nonumber  \\
C_{22} = -\frac{r^{2}a\overset{ .}{a}\overset{ .}{v}}{v} \nonumber  \\
C_{33} = -\frac{r^{2}a\overset{ .}{a}\overset{ .}{v}\sin ^{2}\theta }{v} \nonumber \end{gather}

The tensor $D_{\mu \nu } \equiv \omega \left (\frac{1}{2}g_{\mu \nu } \partial _{\sigma }v \partial ^{\sigma }v - \partial _{\mu }v \partial _{\nu }v\right )$ is computed as:

\begin{align}D_{00} = -\frac{1}{2}\omega \overset{ .}{v}^{2} \label{eq6.6} \\
D_{11} = -\frac{1}{2}\frac{\omega a^{2}\overset{ .}{v}^{2}}{\left (1 -\kappa r^{2}\right )} \nonumber  \\
D_{22} = -\frac{1}{2}r^{2}\omega a^{2}\overset{ .}{v}^{2} \nonumber  \\
D_{33} = -\frac{1}{2}r^{2}\omega a^{2}\overset{ .}{v}^{2}\sin ^{2}\theta  \nonumber \end{align}
while the simpler tensor $E_{\mu \nu } \equiv \omega \left . \partial _{\mu }v \partial _{\nu }v\right .$ has only one non-zero component:
\begin{align}E_{00} =\omega \overset{ .}{v}^{2} \label{eq6.7}\end{align}

From Eqs. (\ref{eq4.13})-(\ref{eq4.14}), the components of the energy-momentum tensor are:

\begin{gather}T_{00} =\rho \left (t\right ) \label{eq6.8} \\
T_{11} =\frac{a^{2}p\left (t\right )}{1 -\kappa r^{2}} \nonumber  \\
T_{22} =r^{2}a^{2}p\left (t\right ) \nonumber  \\
T_{33} =r^{2}a^{2}p\left (t\right )\sin ^{2}\theta  \nonumber \end{gather}with the trace given as $T =T_{\ \mu }^{\mu } = -\rho \left (t\right ) +3p\left (t\right )$.

Using all the above tensor components, the field equation (\ref{eq4.5}) can be written as:
\begin{equation}R_{\mu \nu } -\frac{1}{2}g_{\mu \nu }\left .R\right . +A_{\mu \nu } +B_{\mu \nu } -C_{\mu \nu } +D_{\mu \nu } =8\pi GT_{\mu \nu } \label{eq6.9}
\end{equation}
while the alternative field equation (\ref{eq4.9}) can be computed as:
\begin{equation}R_{\mu \nu } =8\pi G\left (T_{\mu \nu } -\frac{1}{2}g_{\mu \nu }T\right ) +A_{\mu \nu } +\frac{1}{2}B_{\mu \nu } +C_{\mu \nu } +E_{\mu \nu } \label{eq6.10}
\end{equation}

It is usually easier to use this alternative field equation to derive the Friedmann equations. The $\mu \nu  =00$ equation from (\ref{eq6.10}), after some algebraic simplification, gives:

\begin{equation} -3\frac{\overset{ . .}{a}}{a} -\frac{3}{2}\frac{\overset{ .}{a}\overset{ .}{v}}{av} -\frac{3}{2}\frac{\overset{ . .}{v}}{v} -\omega \overset{ .}{v}^{2} +\frac{1}{2}U\left (v\right ) =4\pi G\left (\rho  +3p\right ) \label{eq6.11}
\end{equation}
while the $\mu \nu  =ii$ equations ($i =1 ,2 ,3$) from (\ref{eq6.10}) are all equivalent to each other and yield:

\begin{equation}\frac{\overset{ . .}{a}}{a} +2\genfrac{(}{)}{}{}{\overset{ .}{a}}{a}^{2} +2\frac{\kappa }{a^{2}} +\frac{5}{2}\frac{\overset{ .}{a}\overset{ .}{v}}{av} +\frac{1}{2}\frac{\overset{ . .}{v}}{v} -\frac{1}{2}U\left (v\right ) =4\pi G\left (\rho  -p\right ) . \label{eq6.12}
\end{equation}
Combining these last two equations together, after some simplifications, we obtain the modified Friedmann equations (\ref{eq4.17}) introduced in Sect. \ref{subsect:relativistictwo}.

\bibliographystyle{apsrev4-1}
\bibliography{RFDGmainNotes}

\begin{thebibliography}{54}%
\makeatletter
\providecommand \@ifxundefined [1]{%
 \@ifx{#1\undefined}
}%
\providecommand \@ifnum [1]{%
 \ifnum #1\expandafter \@firstoftwo
 \else \expandafter \@secondoftwo
 \fi
}%
\providecommand \@ifx [1]{%
 \ifx #1\expandafter \@firstoftwo
 \else \expandafter \@secondoftwo
 \fi
}%
\providecommand \natexlab [1]{#1}%
\providecommand \enquote  [1]{``#1''}%
\providecommand \bibnamefont  [1]{#1}%
\providecommand \bibfnamefont [1]{#1}%
\providecommand \citenamefont [1]{#1}%
\providecommand \href@noop [0]{\@secondoftwo}%
\providecommand \href [0]{\begingroup \@sanitize@url \@href}%
\providecommand \@href[1]{\@@startlink{#1}\@@href}%
\providecommand \@@href[1]{\endgroup#1\@@endlink}%
\providecommand \@sanitize@url [0]{\catcode `\\12\catcode `\$12\catcode
  `\&12\catcode `\#12\catcode `\^12\catcode `\_12\catcode `\%12\relax}%
\providecommand \@@startlink[1]{}%
\providecommand \@@endlink[0]{}%
\providecommand \url  [0]{\begingroup\@sanitize@url \@url }%
\providecommand \@url [1]{\endgroup\@href {#1}{\urlprefix }}%
\providecommand \urlprefix  [0]{URL }%
\providecommand \Eprint [0]{\href }%
\providecommand \doibase [0]{http://dx.doi.org/}%
\providecommand \selectlanguage [0]{\@gobble}%
\providecommand \bibinfo  [0]{\@secondoftwo}%
\providecommand \bibfield  [0]{\@secondoftwo}%
\providecommand \translation [1]{[#1]}%
\providecommand \BibitemOpen [0]{}%
\providecommand \bibitemStop [0]{}%
\providecommand \bibitemNoStop [0]{.\EOS\space}%
\providecommand \EOS [0]{\spacefactor3000\relax}%
\providecommand \BibitemShut  [1]{\csname bibitem#1\endcsname}%
\let\auto@bib@innerbib\@empty
\bibitem [{\citenamefont {Varieschi}(2020{\natexlab{a}})}]{Varieschi:2020ioh}%
  \BibitemOpen
  \bibfield  {author} {\bibinfo {author} {\bibfnamefont {G.~U.}\ \bibnamefont
  {Varieschi}},\ }\href {\doibase 10.1007/s10701-020-00389-7} {\bibfield
  {journal} {\bibinfo  {journal} {Found. Phys.}\ }\textbf {\bibinfo {volume}
  {50}},\ \bibinfo {pages} {1608} (\bibinfo {year} {2020}{\natexlab{a}})},\
  \Eprint {http://arxiv.org/abs/2003.05784} {arXiv:2003.05784 [gr-qc]}
  \BibitemShut {NoStop}%
\bibitem [{\citenamefont {Varieschi}(2021)}]{Varieschi:2020dnd}%
  \BibitemOpen
  \bibfield  {author} {\bibinfo {author} {\bibfnamefont {G.~U.}\ \bibnamefont
  {Varieschi}},\ }\href {\doibase 10.1140/epjp/s13360-021-01165-w} {\bibfield
  {journal} {\bibinfo  {journal} {Eur. Phys. J. Plus}\ }\textbf {\bibinfo
  {volume} {136}},\ \bibinfo {pages} {183} (\bibinfo {year} {2021})},\ \Eprint
  {http://arxiv.org/abs/2008.04737} {arXiv:2008.04737 [gr-qc]} \BibitemShut
  {NoStop}%
\bibitem [{\citenamefont {{Varieschi}}(2021)}]{2021MNRAS.503.1915V}%
  \BibitemOpen
  \bibfield  {author} {\bibinfo {author} {\bibfnamefont {G.~U.}\ \bibnamefont
  {{Varieschi}}},\ }\href {\doibase 10.1093/mnras/stab433} {\bibfield
  {journal} {\bibinfo  {journal} {Mon. Not. Roy. Astron. Soc.}\ }\textbf
  {\bibinfo {volume} {503}},\ \bibinfo {pages} {1915} (\bibinfo {year}
  {2021})},\ \Eprint {http://arxiv.org/abs/2011.04911} {arXiv:2011.04911
  [gr-qc]} \BibitemShut {NoStop}%
\bibitem [{\citenamefont {Varieschi}(2020{\natexlab{b}})}]{Varieschi:webpage}%
  \BibitemOpen
  \bibfield  {author} {\bibinfo {author} {\bibfnamefont {G.~U.}\ \bibnamefont
  {Varieschi}},\ }\href {http://gvarieschi.lmu.build/NFDG2020.html} {\emph
  {\bibinfo {title} {Newtonian Fractional-Dimension Gravity (NFDG)-
  http://gvarieschi.lmu.build/NFDG2020.html}}} (\bibinfo {year}
  {2020}{\natexlab{b}})\BibitemShut {NoStop}%
\bibitem [{\citenamefont {Lelli}\ \emph {et~al.}(2016)\citenamefont {Lelli},
  \citenamefont {McGaugh},\ and\ \citenamefont {Schombert}}]{Lelli:2016zqa}%
  \BibitemOpen
  \bibfield  {author} {\bibinfo {author} {\bibfnamefont {F.}~\bibnamefont
  {Lelli}}, \bibinfo {author} {\bibfnamefont {S.~S.}\ \bibnamefont {McGaugh}},
  \ and\ \bibinfo {author} {\bibfnamefont {J.~M.}\ \bibnamefont {Schombert}},\
  }\href {\doibase 10.3847/0004-6256/152/6/157} {\bibfield  {journal} {\bibinfo
   {journal} {Astron. J.}\ }\textbf {\bibinfo {volume} {152}},\ \bibinfo
  {pages} {157} (\bibinfo {year} {2016})},\ \Eprint
  {http://arxiv.org/abs/1606.09251} {arXiv:1606.09251 [astro-ph.GA]}
  \BibitemShut {NoStop}%
\bibitem [{\citenamefont {Varieschi}(2018)}]{Varieschi:2018}%
  \BibitemOpen
  \bibfield  {author} {\bibinfo {author} {\bibfnamefont {G.~U.}\ \bibnamefont
  {Varieschi}},\ }\href {\doibase 10.4236/jamp.2018.66105} {\bibfield
  {journal} {\bibinfo  {journal} {J. Appl. Math.Phys.}\ }\textbf {\bibinfo
  {volume} {06}},\ \bibinfo {pages} {1247} (\bibinfo {year} {2018})},\ \Eprint
  {http://arxiv.org/abs/1712.03473} {arXiv:1712.03473 [physics.class-ph]}
  \BibitemShut {NoStop}%
\bibitem [{\citenamefont {Calcagni}(2010{\natexlab{a}})}]{Calcagni:2009kc}%
  \BibitemOpen
  \bibfield  {author} {\bibinfo {author} {\bibfnamefont {G.}~\bibnamefont
  {Calcagni}},\ }\href {\doibase 10.1103/PhysRevLett.104.251301} {\bibfield
  {journal} {\bibinfo  {journal} {Phys. Rev. Lett.}\ }\textbf {\bibinfo
  {volume} {104}},\ \bibinfo {pages} {251301} (\bibinfo {year}
  {2010}{\natexlab{a}})},\ \Eprint {http://arxiv.org/abs/0912.3142}
  {arXiv:0912.3142 [hep-th]} \BibitemShut {NoStop}%
\bibitem [{\citenamefont {Calcagni}(2010{\natexlab{b}})}]{Calcagni:2010bj}%
  \BibitemOpen
  \bibfield  {author} {\bibinfo {author} {\bibfnamefont {G.}~\bibnamefont
  {Calcagni}},\ }\href {\doibase 10.1007/JHEP03(2010)120} {\bibfield  {journal}
  {\bibinfo  {journal} {JHEP}\ }\textbf {\bibinfo {volume} {03}},\ \bibinfo
  {pages} {120} (\bibinfo {year} {2010}{\natexlab{b}})},\ \Eprint
  {http://arxiv.org/abs/1001.0571} {arXiv:1001.0571 [hep-th]} \BibitemShut
  {NoStop}%
\bibitem [{\citenamefont {Calcagni}(2012{\natexlab{a}})}]{Calcagni:2011kn}%
  \BibitemOpen
  \bibfield  {author} {\bibinfo {author} {\bibfnamefont {G.}~\bibnamefont
  {Calcagni}},\ }\href {\doibase 10.4310/ATMP.2012.v16.n2.a5} {\bibfield
  {journal} {\bibinfo  {journal} {Adv. Theor. Math. Phys.}\ }\textbf {\bibinfo
  {volume} {16}},\ \bibinfo {pages} {549} (\bibinfo {year}
  {2012}{\natexlab{a}})},\ \Eprint {http://arxiv.org/abs/1106.5787}
  {arXiv:1106.5787 [hep-th]} \BibitemShut {NoStop}%
\bibitem [{\citenamefont {Calcagni}(2012{\natexlab{b}})}]{Calcagni:2011sz}%
  \BibitemOpen
  \bibfield  {author} {\bibinfo {author} {\bibfnamefont {G.}~\bibnamefont
  {Calcagni}},\ }\href {\doibase 10.1007/JHEP01(2012)065} {\bibfield  {journal}
  {\bibinfo  {journal} {JHEP}\ }\textbf {\bibinfo {volume} {01}},\ \bibinfo
  {pages} {065} (\bibinfo {year} {2012}{\natexlab{b}})},\ \Eprint
  {http://arxiv.org/abs/1107.5041} {arXiv:1107.5041 [hep-th]} \BibitemShut
  {NoStop}%
\bibitem [{\citenamefont {Calcagni}(2013)}]{Calcagni:2013yqa}%
  \BibitemOpen
  \bibfield  {author} {\bibinfo {author} {\bibfnamefont {G.}~\bibnamefont
  {Calcagni}},\ }\href {\doibase 10.1088/1475-7516/2013/12/041} {\bibfield
  {journal} {\bibinfo  {journal} {JCAP}\ }\textbf {\bibinfo {volume} {12}},\
  \bibinfo {pages} {041} (\bibinfo {year} {2013})},\ \Eprint
  {http://arxiv.org/abs/1307.6382} {arXiv:1307.6382 [hep-th]} \BibitemShut
  {NoStop}%
\bibitem [{\citenamefont {Calcagni}(2017{\natexlab{a}})}]{Calcagni:2016azd}%
  \BibitemOpen
  \bibfield  {author} {\bibinfo {author} {\bibfnamefont {G.}~\bibnamefont
  {Calcagni}},\ }\href {\doibase 10.1007/JHEP03(2017)138} {\bibfield  {journal}
  {\bibinfo  {journal} {JHEP}\ }\textbf {\bibinfo {volume} {03}},\ \bibinfo
  {pages} {138} (\bibinfo {year} {2017}{\natexlab{a}})},\ \bibinfo {note}
  {[Erratum: JHEP 06, 020 (2017)]},\ \Eprint {http://arxiv.org/abs/1612.05632}
  {arXiv:1612.05632 [hep-th]} \BibitemShut {NoStop}%
\bibitem [{\citenamefont {Calcagni}(2017{\natexlab{b}})}]{Calcagni:2016xtk}%
  \BibitemOpen
  \bibfield  {author} {\bibinfo {author} {\bibfnamefont {G.}~\bibnamefont
  {Calcagni}},\ }\href {\doibase 10.1103/PhysRevD.95.064057} {\bibfield
  {journal} {\bibinfo  {journal} {Phys. Rev. D}\ }\textbf {\bibinfo {volume}
  {95}},\ \bibinfo {pages} {064057} (\bibinfo {year} {2017}{\natexlab{b}})},\
  \Eprint {http://arxiv.org/abs/1609.02776} {arXiv:1609.02776 [gr-qc]}
  \BibitemShut {NoStop}%
\bibitem [{\citenamefont {Calcagni}(2018)}]{Calcagni:2018dhp}%
  \BibitemOpen
  \bibfield  {author} {\bibinfo {author} {\bibfnamefont {G.}~\bibnamefont
  {Calcagni}},\ }\href {\doibase 10.3389/fphy.2018.00058} {\bibfield  {journal}
  {\bibinfo  {journal} {Front.in Phys.}\ }\textbf {\bibinfo {volume} {6}},\
  \bibinfo {pages} {58} (\bibinfo {year} {2018})},\ \Eprint
  {http://arxiv.org/abs/1801.00396} {arXiv:1801.00396 [math-ph]} \BibitemShut
  {NoStop}%
\bibitem [{\citenamefont {Calcagni}\ and\ \citenamefont
  {De~Felice}(2020)}]{Calcagni:2020ads}%
  \BibitemOpen
  \bibfield  {author} {\bibinfo {author} {\bibfnamefont {G.}~\bibnamefont
  {Calcagni}}\ and\ \bibinfo {author} {\bibfnamefont {A.}~\bibnamefont
  {De~Felice}},\ }\href {\doibase 10.1103/PhysRevD.102.103529} {\bibfield
  {journal} {\bibinfo  {journal} {Phys. Rev. D}\ }\textbf {\bibinfo {volume}
  {102}},\ \bibinfo {pages} {103529} (\bibinfo {year} {2020})},\ \Eprint
  {http://arxiv.org/abs/2004.02896} {arXiv:2004.02896 [gr-qc]} \BibitemShut
  {NoStop}%
\bibitem [{\citenamefont {Calcagni}(2021{\natexlab{a}})}]{Calcagni:2021ljs}%
  \BibitemOpen
  \bibfield  {author} {\bibinfo {author} {\bibfnamefont {G.}~\bibnamefont
  {Calcagni}},\ }\href {\doibase 10.1088/1361-6382/ac103c} {\bibfield
  {journal} {\bibinfo  {journal} {Class. Quant. Grav.}\ }\textbf {\bibinfo
  {volume} {38}},\ \bibinfo {pages} {165006} (\bibinfo {year}
  {2021}{\natexlab{a}})},\ \Eprint {http://arxiv.org/abs/2102.03363}
  {arXiv:2102.03363 [hep-th]} \BibitemShut {NoStop}%
\bibitem [{\citenamefont {Calcagni}(2021{\natexlab{b}})}]{Calcagni:2021ipd}%
  \BibitemOpen
  \bibfield  {author} {\bibinfo {author} {\bibfnamefont {G.}~\bibnamefont
  {Calcagni}},\ }\href {\doibase 10.1142/S021773232140006X} {\bibfield
  {journal} {\bibinfo  {journal} {Mod. Phys. Lett. A}\ }\textbf {\bibinfo
  {volume} {36}},\ \bibinfo {pages} {2140006} (\bibinfo {year}
  {2021}{\natexlab{b}})},\ \Eprint {http://arxiv.org/abs/2103.06557}
  {arXiv:2103.06557 [hep-th]} \BibitemShut {NoStop}%
\bibitem [{\citenamefont {Calcagni}(2021{\natexlab{c}})}]{Calcagni:2021aap}%
  \BibitemOpen
  \bibfield  {author} {\bibinfo {author} {\bibfnamefont {G.}~\bibnamefont
  {Calcagni}},\ }\href {\doibase 10.1088/1361-6382/ac1bea} {\bibfield
  {journal} {\bibinfo  {journal} {Class. Quant. Grav.}\ }\textbf {\bibinfo
  {volume} {38}},\ \bibinfo {pages} {165005} (\bibinfo {year}
  {2021}{\natexlab{c}})},\ \bibinfo {note} {[Erratum: Class.Quant.Grav. 38,
  169601 (2021)]},\ \Eprint {http://arxiv.org/abs/2106.15430} {arXiv:2106.15430
  [gr-qc]} \BibitemShut {NoStop}%
\bibitem [{\citenamefont {Giusti}(2020)}]{Giusti:2020rul}%
  \BibitemOpen
  \bibfield  {author} {\bibinfo {author} {\bibfnamefont {A.}~\bibnamefont
  {Giusti}},\ }\href {\doibase 10.1103/PhysRevD.101.124029} {\bibfield
  {journal} {\bibinfo  {journal} {Phys. Rev. D}\ }\textbf {\bibinfo {volume}
  {101}},\ \bibinfo {pages} {124029} (\bibinfo {year} {2020})},\ \Eprint
  {http://arxiv.org/abs/2002.07133} {arXiv:2002.07133 [gr-qc]} \BibitemShut
  {NoStop}%
\bibitem [{\citenamefont {Giusti}\ \emph {et~al.}(2020)\citenamefont {Giusti},
  \citenamefont {Garrappa},\ and\ \citenamefont {Vachon}}]{Giusti:2020kcv}%
  \BibitemOpen
  \bibfield  {author} {\bibinfo {author} {\bibfnamefont {A.}~\bibnamefont
  {Giusti}}, \bibinfo {author} {\bibfnamefont {R.}~\bibnamefont {Garrappa}}, \
  and\ \bibinfo {author} {\bibfnamefont {G.}~\bibnamefont {Vachon}},\ }\href
  {\doibase 10.1140/epjp/s13360-020-00831-9} {\bibfield  {journal} {\bibinfo
  {journal} {Eur. Phys. J. Plus}\ }\textbf {\bibinfo {volume} {135}},\ \bibinfo
  {pages} {798} (\bibinfo {year} {2020})},\ \Eprint
  {http://arxiv.org/abs/2009.04335} {arXiv:2009.04335 [gr-qc]} \BibitemShut
  {NoStop}%
\bibitem [{\citenamefont {Milgrom}(1983{\natexlab{a}})}]{Milgrom:1983ca}%
  \BibitemOpen
  \bibfield  {author} {\bibinfo {author} {\bibfnamefont {M.}~\bibnamefont
  {Milgrom}},\ }\href {\doibase 10.1086/161130} {\bibfield  {journal} {\bibinfo
   {journal} {Astrophys. J.}\ }\textbf {\bibinfo {volume} {270}},\ \bibinfo
  {pages} {365} (\bibinfo {year} {1983}{\natexlab{a}})}\BibitemShut {NoStop}%
\bibitem [{\citenamefont {Milgrom}(1983{\natexlab{b}})}]{Milgrom:1983pn}%
  \BibitemOpen
  \bibfield  {author} {\bibinfo {author} {\bibfnamefont {M.}~\bibnamefont
  {Milgrom}},\ }\href {\doibase 10.1086/161131} {\bibfield  {journal} {\bibinfo
   {journal} {Astrophys. J.}\ }\textbf {\bibinfo {volume} {270}},\ \bibinfo
  {pages} {371} (\bibinfo {year} {1983}{\natexlab{b}})}\BibitemShut {NoStop}%
\bibitem [{\citenamefont {Milgrom}(1983{\natexlab{c}})}]{Milgrom:1983zz}%
  \BibitemOpen
  \bibfield  {author} {\bibinfo {author} {\bibfnamefont {M.}~\bibnamefont
  {Milgrom}},\ }\href {\doibase 10.1086/161132} {\bibfield  {journal} {\bibinfo
   {journal} {Astrophys. J.}\ }\textbf {\bibinfo {volume} {270}},\ \bibinfo
  {pages} {384} (\bibinfo {year} {1983}{\natexlab{c}})}\BibitemShut {NoStop}%
\bibitem [{\citenamefont {McGaugh}\ \emph {et~al.}(2016)\citenamefont
  {McGaugh}, \citenamefont {Lelli},\ and\ \citenamefont
  {Schombert}}]{McGaugh:2016leg}%
  \BibitemOpen
  \bibfield  {author} {\bibinfo {author} {\bibfnamefont {S.}~\bibnamefont
  {McGaugh}}, \bibinfo {author} {\bibfnamefont {F.}~\bibnamefont {Lelli}}, \
  and\ \bibinfo {author} {\bibfnamefont {J.}~\bibnamefont {Schombert}},\ }\href
  {\doibase 10.1103/PhysRevLett.117.201101} {\bibfield  {journal} {\bibinfo
  {journal} {Phys. Rev. Lett.}\ }\textbf {\bibinfo {volume} {117}},\ \bibinfo
  {pages} {201101} (\bibinfo {year} {2016})},\ \Eprint
  {http://arxiv.org/abs/1609.05917} {arXiv:1609.05917 [astro-ph.GA]}
  \BibitemShut {NoStop}%
\bibitem [{\citenamefont {Lelli}\ \emph {et~al.}(2017)\citenamefont {Lelli},
  \citenamefont {McGaugh}, \citenamefont {Schombert},\ and\ \citenamefont
  {Pawlowski}}]{Lelli:2017vgz}%
  \BibitemOpen
  \bibfield  {author} {\bibinfo {author} {\bibfnamefont {F.}~\bibnamefont
  {Lelli}}, \bibinfo {author} {\bibfnamefont {S.~S.}\ \bibnamefont {McGaugh}},
  \bibinfo {author} {\bibfnamefont {J.~M.}\ \bibnamefont {Schombert}}, \ and\
  \bibinfo {author} {\bibfnamefont {M.~S.}\ \bibnamefont {Pawlowski}},\ }\href
  {\doibase 10.3847/1538-4357/836/2/152} {\bibfield  {journal} {\bibinfo
  {journal} {Astrophys. J.}\ }\textbf {\bibinfo {volume} {836}},\ \bibinfo
  {pages} {152} (\bibinfo {year} {2017})},\ \Eprint
  {http://arxiv.org/abs/1610.08981} {arXiv:1610.08981 [astro-ph.GA]}
  \BibitemShut {NoStop}%
\bibitem [{\citenamefont {Chae}\ \emph {et~al.}(2020)\citenamefont {Chae},
  \citenamefont {Lelli}, \citenamefont {Desmond}, \citenamefont {McGaugh},
  \citenamefont {Li},\ and\ \citenamefont {Schombert}}]{Chae:2020omu}%
  \BibitemOpen
  \bibfield  {author} {\bibinfo {author} {\bibfnamefont {K.-H.}\ \bibnamefont
  {Chae}}, \bibinfo {author} {\bibfnamefont {F.}~\bibnamefont {Lelli}},
  \bibinfo {author} {\bibfnamefont {H.}~\bibnamefont {Desmond}}, \bibinfo
  {author} {\bibfnamefont {S.~S.}\ \bibnamefont {McGaugh}}, \bibinfo {author}
  {\bibfnamefont {P.}~\bibnamefont {Li}}, \ and\ \bibinfo {author}
  {\bibfnamefont {J.~M.}\ \bibnamefont {Schombert}},\ }\href {\doibase
  10.3847/1538-4357/abbb96} {\bibfield  {journal} {\bibinfo  {journal}
  {Astrophys. J.}\ }\textbf {\bibinfo {volume} {904}},\ \bibinfo {pages} {51}
  (\bibinfo {year} {2020})},\ \bibinfo {note} {[Erratum: Astrophys.J. 910, 81
  (2021)]},\ \Eprint {http://arxiv.org/abs/2009.11525} {arXiv:2009.11525
  [astro-ph.GA]} \BibitemShut {NoStop}%
\bibitem [{\citenamefont {{Sadallah}}\ \emph {et~al.}(2006)\citenamefont
  {{Sadallah}}, \citenamefont {{Muslih}},\ and\ \citenamefont
  {{Baleanu}}}]{2006CzJPh..56..323S}%
  \BibitemOpen
  \bibfield  {author} {\bibinfo {author} {\bibfnamefont {M.}~\bibnamefont
  {{Sadallah}}}, \bibinfo {author} {\bibfnamefont {S.~I.}\ \bibnamefont
  {{Muslih}}}, \ and\ \bibinfo {author} {\bibfnamefont {D.}~\bibnamefont
  {{Baleanu}}},\ }\href {\doibase 10.1007/s10582-006-0093-7} {\bibfield
  {journal} {\bibinfo  {journal} {Czechoslovak Journal of Physics}\ }\textbf
  {\bibinfo {volume} {56}},\ \bibinfo {pages} {323} (\bibinfo {year}
  {2006})}\BibitemShut {NoStop}%
\bibitem [{\citenamefont {Sadallah}\ and\ \citenamefont
  {Muslih}(2009)}]{Sadallah:2009zz}%
  \BibitemOpen
  \bibfield  {author} {\bibinfo {author} {\bibfnamefont {M.}~\bibnamefont
  {Sadallah}}\ and\ \bibinfo {author} {\bibfnamefont {S.~I.}\ \bibnamefont
  {Muslih}},\ }\href {\doibase 10.1007/s10773-009-0133-8} {\bibfield  {journal}
  {\bibinfo  {journal} {Int. J. Theor. Phys.}\ }\textbf {\bibinfo {volume}
  {48}},\ \bibinfo {pages} {3312} (\bibinfo {year} {2009})}\BibitemShut
  {NoStop}%
\bibitem [{\citenamefont {{Collas}}(1977)}]{1977AmJPh..45..833C}%
  \BibitemOpen
  \bibfield  {author} {\bibinfo {author} {\bibfnamefont {P.}~\bibnamefont
  {{Collas}}},\ }\href {\doibase 10.1119/1.11057} {\bibfield  {journal}
  {\bibinfo  {journal} {American Journal of Physics}\ }\textbf {\bibinfo
  {volume} {45}},\ \bibinfo {pages} {833} (\bibinfo {year} {1977})}\BibitemShut
  {NoStop}%
\bibitem [{\citenamefont {Romero}\ and\ \citenamefont
  {Dahia}(1994)}]{Romero:1994va}%
  \BibitemOpen
  \bibfield  {author} {\bibinfo {author} {\bibfnamefont {C.}~\bibnamefont
  {Romero}}\ and\ \bibinfo {author} {\bibfnamefont {F.}~\bibnamefont {Dahia}},\
  }\href {\doibase 10.1007/BF00675174} {\bibfield  {journal} {\bibinfo
  {journal} {Int. J. Theor. Phys.}\ }\textbf {\bibinfo {volume} {33}},\
  \bibinfo {pages} {2091} (\bibinfo {year} {1994})}\BibitemShut {NoStop}%
\bibitem [{\citenamefont {{Deser}}\ \emph {et~al.}(1984)\citenamefont
  {{Deser}}, \citenamefont {{Jackiw}},\ and\ \citenamefont {{'t
  Hooft}}}]{1984AnPhy.152..220D}%
  \BibitemOpen
  \bibfield  {author} {\bibinfo {author} {\bibfnamefont {S.}~\bibnamefont
  {{Deser}}}, \bibinfo {author} {\bibfnamefont {R.}~\bibnamefont {{Jackiw}}}, \
  and\ \bibinfo {author} {\bibfnamefont {G.}~\bibnamefont {{'t Hooft}}},\
  }\href {\doibase 10.1016/0003-4916(84)90085-X} {\bibfield  {journal}
  {\bibinfo  {journal} {Annals of Physics}\ }\textbf {\bibinfo {volume}
  {152}},\ \bibinfo {pages} {220} (\bibinfo {year} {1984})}\BibitemShut
  {NoStop}%
\bibitem [{\citenamefont {Clifton}\ \emph {et~al.}(2012)\citenamefont
  {Clifton}, \citenamefont {Ferreira}, \citenamefont {Padilla},\ and\
  \citenamefont {Skordis}}]{CLIFTON20121}%
  \BibitemOpen
  \bibfield  {author} {\bibinfo {author} {\bibfnamefont {T.}~\bibnamefont
  {Clifton}}, \bibinfo {author} {\bibfnamefont {P.~G.}\ \bibnamefont
  {Ferreira}}, \bibinfo {author} {\bibfnamefont {A.}~\bibnamefont {Padilla}}, \
  and\ \bibinfo {author} {\bibfnamefont {C.}~\bibnamefont {Skordis}},\ }\href
  {\doibase https://doi.org/10.1016/j.physrep.2012.01.001} {\bibfield
  {journal} {\bibinfo  {journal} {Physics Reports}\ }\textbf {\bibinfo {volume}
  {513}},\ \bibinfo {pages} {1} (\bibinfo {year} {2012})},\ \bibinfo {note}
  {modified Gravity and Cosmology}\BibitemShut {NoStop}%
\bibitem [{\citenamefont {Saridakis}\ \emph {et~al.}(2021)\citenamefont
  {Saridakis} \emph {et~al.}}]{CANTATA:2021ktz}%
  \BibitemOpen
  \bibfield  {author} {\bibinfo {author} {\bibfnamefont {E.~N.}\ \bibnamefont
  {Saridakis}} \emph {et~al.} (\bibinfo {collaboration} {CANTATA}),\
  }\href@noop {} {\  (\bibinfo {year} {2021})},\ \Eprint
  {http://arxiv.org/abs/2105.12582} {arXiv:2105.12582 [gr-qc]} \BibitemShut
  {NoStop}%
\bibitem [{\citenamefont {Will}(2014)}]{Will:2014kxa}%
  \BibitemOpen
  \bibfield  {author} {\bibinfo {author} {\bibfnamefont {C.~M.}\ \bibnamefont
  {Will}},\ }\href {\doibase 10.12942/lrr-2014-4} {\bibfield  {journal}
  {\bibinfo  {journal} {Living Rev. Rel.}\ }\textbf {\bibinfo {volume} {17}},\
  \bibinfo {pages} {4} (\bibinfo {year} {2014})},\ \Eprint
  {http://arxiv.org/abs/1403.7377} {arXiv:1403.7377 [gr-qc]} \BibitemShut
  {NoStop}%
\bibitem [{\citenamefont {Sanders}\ and\ \citenamefont
  {McGaugh}(2002)}]{Sanders:2002pf}%
  \BibitemOpen
  \bibfield  {author} {\bibinfo {author} {\bibfnamefont {R.~H.}\ \bibnamefont
  {Sanders}}\ and\ \bibinfo {author} {\bibfnamefont {S.~S.}\ \bibnamefont
  {McGaugh}},\ }\href {\doibase 10.1146/annurev.astro.40.060401.093923}
  {\bibfield  {journal} {\bibinfo  {journal} {Ann. Rev. Astron. Astrophys.}\
  }\textbf {\bibinfo {volume} {40}},\ \bibinfo {pages} {263} (\bibinfo {year}
  {2002})},\ \Eprint {http://arxiv.org/abs/astro-ph/0204521}
  {arXiv:astro-ph/0204521} \BibitemShut {NoStop}%
\bibitem [{\citenamefont {Famaey}\ and\ \citenamefont
  {McGaugh}(2012)}]{Famaey:2011kh}%
  \BibitemOpen
  \bibfield  {author} {\bibinfo {author} {\bibfnamefont {B.}~\bibnamefont
  {Famaey}}\ and\ \bibinfo {author} {\bibfnamefont {S.}~\bibnamefont
  {McGaugh}},\ }\href {\doibase 10.12942/lrr-2012-10} {\bibfield  {journal}
  {\bibinfo  {journal} {Living Rev. Rel.}\ }\textbf {\bibinfo {volume} {15}},\
  \bibinfo {pages} {10} (\bibinfo {year} {2012})},\ \Eprint
  {http://arxiv.org/abs/1112.3960} {arXiv:1112.3960 [astro-ph.CO]} \BibitemShut
  {NoStop}%
\bibitem [{\citenamefont {Barrow}(1983)}]{10.2307/37418}%
  \BibitemOpen
  \bibfield  {author} {\bibinfo {author} {\bibfnamefont {J.~D.}\ \bibnamefont
  {Barrow}},\ }\href {http://www.jstor.org/stable/37418} {\bibfield  {journal}
  {\bibinfo  {journal} {Philosophical Transactions of the Royal Society of
  London. Series A, Mathematical and Physical Sciences}\ }\textbf {\bibinfo
  {volume} {310}},\ \bibinfo {pages} {337} (\bibinfo {year}
  {1983})}\BibitemShut {NoStop}%
\bibitem [{\citenamefont {Ehrenfest}(1920)}]{doi:10.1002/andp.19203660503}%
  \BibitemOpen
  \bibfield  {author} {\bibinfo {author} {\bibfnamefont {P.}~\bibnamefont
  {Ehrenfest}},\ }\href {\doibase 10.1002/andp.19203660503} {\bibfield
  {journal} {\bibinfo  {journal} {Annalen der Physik}\ }\textbf {\bibinfo
  {volume} {366}},\ \bibinfo {pages} {440} (\bibinfo {year} {1920})},\ \Eprint
  {http://arxiv.org/abs/https://onlinelibrary.wiley.com/doi/pdf/10.1002/andp.19203660503}
  {https://onlinelibrary.wiley.com/doi/pdf/10.1002/andp.19203660503}
  \BibitemShut {NoStop}%
\bibitem [{\citenamefont {{Callender}}(2005)}]{2005SHPMP..36..113C}%
  \BibitemOpen
  \bibfield  {author} {\bibinfo {author} {\bibfnamefont {C.}~\bibnamefont
  {{Callender}}},\ }\href {\doibase 10.1016/j.shpsb.2004.09.002} {\bibfield
  {journal} {\bibinfo  {journal} {Studies in the History and Philosophy of
  Modern Physics}\ }\textbf {\bibinfo {volume} {36}},\ \bibinfo {pages} {113}
  (\bibinfo {year} {2005})}\BibitemShut {NoStop}%
\bibitem [{\citenamefont {Bollini}\ and\ \citenamefont
  {Giambiagi}(1972)}]{Bollini:1972ui}%
  \BibitemOpen
  \bibfield  {author} {\bibinfo {author} {\bibfnamefont {C.~G.}\ \bibnamefont
  {Bollini}}\ and\ \bibinfo {author} {\bibfnamefont {J.~J.}\ \bibnamefont
  {Giambiagi}},\ }\href {\doibase 10.1007/BF02895558} {\bibfield  {journal}
  {\bibinfo  {journal} {Nuovo Cim.}\ }\textbf {\bibinfo {volume} {B12}},\
  \bibinfo {pages} {20} (\bibinfo {year} {1972})}\BibitemShut {NoStop}%
\bibitem [{\citenamefont {'t~Hooft}\ and\ \citenamefont
  {Veltman}(1972)}]{tHooft:1972tcz}%
  \BibitemOpen
  \bibfield  {author} {\bibinfo {author} {\bibfnamefont {G.}~\bibnamefont
  {'t~Hooft}}\ and\ \bibinfo {author} {\bibfnamefont {M.~J.~G.}\ \bibnamefont
  {Veltman}},\ }\href {\doibase 10.1016/0550-3213(72)90279-9} {\bibfield
  {journal} {\bibinfo  {journal} {Nucl. Phys.}\ }\textbf {\bibinfo {volume}
  {B44}},\ \bibinfo {pages} {189} (\bibinfo {year} {1972})}\BibitemShut
  {NoStop}%
\bibitem [{\citenamefont {Wilson}(1973)}]{Wilson:1972cf}%
  \BibitemOpen
  \bibfield  {author} {\bibinfo {author} {\bibfnamefont {K.~G.}\ \bibnamefont
  {Wilson}},\ }\href {\doibase 10.1103/PhysRevD.7.2911} {\bibfield  {journal}
  {\bibinfo  {journal} {Phys. Rev.}\ }\textbf {\bibinfo {volume} {D7}},\
  \bibinfo {pages} {2911} (\bibinfo {year} {1973})}\BibitemShut {NoStop}%
\bibitem [{\citenamefont {{Peskin}}\ and\ \citenamefont
  {{Schroeder}}(1995)}]{1995iqft.book.....P}%
  \BibitemOpen
  \bibfield  {author} {\bibinfo {author} {\bibfnamefont {M.~E.}\ \bibnamefont
  {{Peskin}}}\ and\ \bibinfo {author} {\bibfnamefont {D.~V.}\ \bibnamefont
  {{Schroeder}}},\ }\href@noop {} {\emph {\bibinfo {title} {An Introduction to
  Quantum Field Theory, by Michael Edward Peskin and Daniel V. Schroeder. ISBN
  13 978-0-201-50397-5, ISBN-10 0-201-50397-2. Published by Westview Press,
  Perseus Books Group, 1995.}}}\ (\bibinfo {year} {1995})\BibitemShut {NoStop}%
\bibitem [{\citenamefont {Stillinger}(1977)}]{doi:10.1063/1.523395}%
  \BibitemOpen
  \bibfield  {author} {\bibinfo {author} {\bibfnamefont {F.~H.}\ \bibnamefont
  {Stillinger}},\ }\href {\doibase 10.1063/1.523395} {\bibfield  {journal}
  {\bibinfo  {journal} {Journal of Mathematical Physics}\ }\textbf {\bibinfo
  {volume} {18}},\ \bibinfo {pages} {1224} (\bibinfo {year} {1977})},\ \Eprint
  {http://arxiv.org/abs/https://doi.org/10.1063/1.523395}
  {https://doi.org/10.1063/1.523395} \BibitemShut {NoStop}%
\bibitem [{\citenamefont {{Svozil}}(1987)}]{1987JPhA...20.3861S}%
  \BibitemOpen
  \bibfield  {author} {\bibinfo {author} {\bibfnamefont {K.}~\bibnamefont
  {{Svozil}}},\ }\href {\doibase 10.1088/0305-4470/20/12/033} {\bibfield
  {journal} {\bibinfo  {journal} {Journal of Physics A Mathematical General}\
  }\textbf {\bibinfo {volume} {20}},\ \bibinfo {pages} {3861} (\bibinfo {year}
  {1987})}\BibitemShut {NoStop}%
\bibitem [{\citenamefont {Palmer}\ and\ \citenamefont
  {Stavrinou}(2004)}]{Palmer_2004}%
  \BibitemOpen
  \bibfield  {author} {\bibinfo {author} {\bibfnamefont {C.}~\bibnamefont
  {Palmer}}\ and\ \bibinfo {author} {\bibfnamefont {P.~N.}\ \bibnamefont
  {Stavrinou}},\ }\href {\doibase 10.1088/0305-4470/37/27/009} {\bibfield
  {journal} {\bibinfo  {journal} {Journal of Physics A: Mathematical and
  General}\ }\textbf {\bibinfo {volume} {37}},\ \bibinfo {pages} {6987}
  (\bibinfo {year} {2004})}\BibitemShut {NoStop}%
\bibitem [{\citenamefont {Tarasov}(2011)}]{bookTarasov}%
  \BibitemOpen
  \bibfield  {author} {\bibinfo {author} {\bibfnamefont {V.}~\bibnamefont
  {Tarasov}},\ }\href@noop {} {\emph {\bibinfo {title} {Fractional Dynamics:
  Application of Fractional Calculus to Dynamics of Particles, Fields and
  Media}}}\ (\bibinfo {year} {2011})\BibitemShut {NoStop}%
\bibitem [{\citenamefont {Zubair}\ \emph {et~al.}(2012)\citenamefont {Zubair},
  \citenamefont {Mughal},\ and\ \citenamefont {Naqvi}}]{bookZubair}%
  \BibitemOpen
  \bibfield  {author} {\bibinfo {author} {\bibfnamefont {M.}~\bibnamefont
  {Zubair}}, \bibinfo {author} {\bibfnamefont {M.}~\bibnamefont {Mughal}}, \
  and\ \bibinfo {author} {\bibfnamefont {Q.}~\bibnamefont {Naqvi}},\ }\href
  {\doibase 10.1007/978-3-642-25358-4} {\emph {\bibinfo {title}
  {Electromagnetic Fields and Waves in Fractional Dimensional Space}}}\
  (\bibinfo {year} {2012})\BibitemShut {NoStop}%
\bibitem [{\citenamefont {Tarasov}(2014)}]{Tarasov:2014fda}%
  \BibitemOpen
  \bibfield  {author} {\bibinfo {author} {\bibfnamefont {V.~E.}\ \bibnamefont
  {Tarasov}},\ }\href {\doibase 10.1063/1.4892155} {\bibfield  {journal}
  {\bibinfo  {journal} {J. Math. Phys.}\ }\textbf {\bibinfo {volume} {55}},\
  \bibinfo {pages} {083510} (\bibinfo {year} {2014})},\ \Eprint
  {http://arxiv.org/abs/1503.02392} {arXiv:1503.02392 [math-ph]} \BibitemShut
  {NoStop}%
\bibitem [{\citenamefont {Tarasov}(2015)}]{TARASOV2015360}%
  \BibitemOpen
  \bibfield  {author} {\bibinfo {author} {\bibfnamefont {V.~E.}\ \bibnamefont
  {Tarasov}},\ }\href {\doibase https://doi.org/10.1016/j.cnsns.2014.05.025}
  {\bibfield  {journal} {\bibinfo  {journal} {Communications in Nonlinear
  Science and Numerical Simulation}\ }\textbf {\bibinfo {volume} {20}},\
  \bibinfo {pages} {360 } (\bibinfo {year} {2015})}\BibitemShut {NoStop}%
\bibitem [{\citenamefont {Morse}\ and\ \citenamefont
  {Feshbach}(1953)}]{morse1953methods}%
  \BibitemOpen
  \bibfield  {author} {\bibinfo {author} {\bibfnamefont {P.}~\bibnamefont
  {Morse}}\ and\ \bibinfo {author} {\bibfnamefont {H.}~\bibnamefont
  {Feshbach}},\ }\href {https://books.google.com/books?id=l8ENAQAAIAAJ} {\emph
  {\bibinfo {title} {Methods of theoretical physics}}},\ International series
  in pure and applied physics\ (\bibinfo  {publisher} {McGraw-Hill},\ \bibinfo
  {year} {1953})\BibitemShut {NoStop}%
\bibitem [{\citenamefont {Carroll}(2019)}]{Carroll:2004st}%
  \BibitemOpen
  \bibfield  {author} {\bibinfo {author} {\bibfnamefont {S.~M.}\ \bibnamefont
  {Carroll}},\ }\href@noop {} {\emph {\bibinfo {title} {{Spacetime and
  Geometry}}}}\ (\bibinfo  {publisher} {Cambridge University Press},\ \bibinfo
  {year} {2019})\BibitemShut {NoStop}%
\bibitem [{\citenamefont {Tsujikawa}(2013)}]{Tsujikawa:2013fta}%
  \BibitemOpen
  \bibfield  {author} {\bibinfo {author} {\bibfnamefont {S.}~\bibnamefont
  {Tsujikawa}},\ }\href {\doibase 10.1088/0264-9381/30/21/214003} {\bibfield
  {journal} {\bibinfo  {journal} {Class. Quant. Grav.}\ }\textbf {\bibinfo
  {volume} {30}},\ \bibinfo {pages} {214003} (\bibinfo {year} {2013})},\
  \Eprint {http://arxiv.org/abs/1304.1961} {arXiv:1304.1961 [gr-qc]}
  \BibitemShut {NoStop}%
\bibitem [{\citenamefont {Carroll}(2001)}]{Carroll:2000fy}%
  \BibitemOpen
  \bibfield  {author} {\bibinfo {author} {\bibfnamefont {S.~M.}\ \bibnamefont
  {Carroll}},\ }\href {\doibase 10.12942/lrr-2001-1} {\bibfield  {journal}
  {\bibinfo  {journal} {Living Rev. Rel.}\ }\textbf {\bibinfo {volume} {4}},\
  \bibinfo {pages} {1} (\bibinfo {year} {2001})},\ \Eprint
  {http://arxiv.org/abs/astro-ph/0004075} {arXiv:astro-ph/0004075} \BibitemShut
  {NoStop}%
\end{thebibliography}%

\end{document}